\documentclass[iop]{emulateapj}

\usepackage{natbib}
\usepackage{multirow}
\usepackage{color}

\usepackage{bm}

\newcommand{\eg}{e.g., }
\newcommand{\ie}{i.e., }
\newcommand{\Msun}{M_{\odot}}

\newcommand{\Mej}{M_{\rm ej}}
\newcommand{\KE}{E_{\rm K}}
\newcommand{\Ye}{Y_{\rm e}}
\def\gsim{\mathrel{\rlap{\lower 4pt \hbox{\hskip 1pt $\sim$}}\raise 1pt
\hbox {$>$}}}
\def\lsim{\mathrel{\rlap{\lower 4pt \hbox{\hskip 1pt $\sim$}}\raise 1pt
\hbox {$<$}}}

\def\ion#1#2{{\rm #1}~{\sc #2}}

\shorttitle{}
\shortauthors{}

\begin{document}

\title{
  Properties of Kilonovae from Dynamical and Post-Merger Ejecta of Neutron Star Mergers
}
\author{
  Masaomi Tanaka\altaffilmark{1},
  Daiji Kato\altaffilmark{2,3,4},
  Gediminas Gaigalas\altaffilmark{5},
  Pavel Rynkun\altaffilmark{5},
  Laima Rad\v{z}i\={u}t\.{e}\altaffilmark{5},
  Shinya Wanajo\altaffilmark{6,7},
  Yuichiro Sekiguchi\altaffilmark{8},
  Nobuyuki Nakamura\altaffilmark{9},
  Hajime Tanuma\altaffilmark{10},
  Izumi Murakami\altaffilmark{2,3},
  Hiroyuki A. Sakaue\altaffilmark{2}
}

\altaffiltext{1}{National Astronomical Observatory of Japan, Osawa, Mitaka, Tokyo 181-8588, Japan; masaomi.tanaka@nao.ac.jp}
\altaffiltext{2}{National Institute for Fusion Science, Oroshi-cho, Toki, Gifu 509-5292, Japan}
\altaffiltext{3}{Department of Fusion Science, SOKENDAI, Oroshi-cho, Toki, Gifu 509-5292, Japan}
\altaffiltext{4}{Department of Advanced Energy Engineering, Kyushu University, Kasuga, Fukuoka 816-8580, Japan}
\altaffiltext{5}{Institute of Theoretical Physics and Astronomy, Vilnius University, Saul\.{e}tekio av. 3, LT-10257 Vilnius, Lithuania}
\altaffiltext{6}{Department of Engineering and Applied Sciences, Sophia University, Chiyoda-ku, Tokyo 102-8554, Japan}
\altaffiltext{7}{RIKEN, iTHES Research Group, 2-1 Hirosawa, Wako, Saitama 351-0198, Japan}
\altaffiltext{8}{Department of Physics, Toho University, Funabashi, Chiba 274-8510, Japan}
\altaffiltext{9}{Institute for Laser Science, The University of Electro-Communications, Chofugaoka, Chofu Tokyo 182-8585, Japan}
\altaffiltext{10}{Department of Physics, Tokyo Metropolitan University, Minami-Osawa, Hachioji, Tokyo 192-0397, Japan}

\begin{abstract}
  Ejected material from neutron star mergers give rise to
  electromagnetic emission powered by radioactive decays of $r$-process nuclei,
  which is so called kilonova or macronova.
  While properties of the emission are largely affected by opacities in the
  ejected material,
  available atomic data for $r$-process elements are still limited.
  We perform atomic structure calculations for
  $r$-process elements:
  Se ($Z=34$), Ru ($Z=44$), Te ($Z=52$), Ba ($Z=56$), Nd ($Z=60$), and Er ($Z=68$).
  We confirm that the opacities from bound-bound transitions of open f-shell,
  Lanthanide elements (Nd and Er) are higher than those of the other elements
  over a wide wavelength range.
  The opacities of open s-shell (Ba), p-shell (Se and Te), and d-shell (Ru)
  elements are lower than those of open f-shell elements and
  their transitions are concentrated in the ultraviolet wavelengths.
  We show that the optical brightness can be different by $> 2$ mag
  depending on the element abundances in the ejecta
  such that post-merger, Lanthanide-free ejecta produce
  brighter and bluer optical emission.
  Such blue emission from post-merger ejecta can be observed from the polar
  directions
  if the mass of the preceding dynamical ejecta in these regions is small.
  For the ejecta mass of 0.01 $\Msun$, observed magnitudes
  of the blue emission will reach
  21.0 mag (100 Mpc) and 22.5 mag (200 Mpc)
  in $g$ and $r$ bands within a few days after the merger,
  which are detectable with 1m or 2m-class telescopes.
\end{abstract}


\section{Introduction}
\label{sec:introduction}

Direct detection of gravitational waves (GW)
opened the era of GW astronomy \citep{abbott16,abbott16b,abbott17}.
A next important step will be an identification of
their electromagnetic (EM) counterparts
to further study the astrophysical nature of the GW sources,
as sky localization by GW detectors is not accurate enough
to pin down their positions \citep{abbott16review}.
In fact, extensive EM follow-up observations have been performed
for the detected GW events so far \citep{abbott16followup}.

From compact binary mergers including at least one neutron star (NS),
\ie NS-NS mergers and black hole (BH)-NS mergers,
various EM signals are expected over a wide wavelength range
\citep[\eg][]{metzger12,rosswog15}.
One of the most promising EM transients is 
so called ``kilonova'' or ``macronova'',
which is the emission powered by the radioactive decays of newly synthesized
$r$-process nuclei \citep{li98,kulkarni05,metzger10}.
For recent reviews of kilonova emission, see
\citet{fernandez16}, \citet{tanaka16}, and \citet{metzger17}.
Kilonova emission is a good candidate for optical and near infrared
follow-up observations after the detection of GWs
\citep{smartt16,soares-santos16,kasliwal16,morokuma16,cowperthwaite16,yoshida17}.

Kilonova emission from the dynamical ejecta ($\sim 10^{-3} - 10^{-2} \Msun$)
of neutron star mergers
is likely to have a luminosity of $\sim 10^{40}-10^{41}\ {\rm erg\ s^{-1}}$
with a timescale of about 1 week, 
which is expected to peak at red optical or near-infrared wavelengths
\citep{kasen13,barnes13,tanaka13}.
This is due to the high opacities of $r$-process elements in the ejecta,
especially those of Lanthanide elements \citep{kasen13}.
In fact, short GRB 130603B showed an near-infrared
excess in the afterglow \citep{tanvir13,berger13},
which was interpreted as a kilonova signal
\citep[see also][]{hotokezaka13b,piran14}.
In addition, possible kilonova candidates have been reported
for GRB 060614 \citep{yang15,jin15} and GRB 050709 \citep{jin16}.

After the dynamical mass ejection,
NS-NS mergers and BH-NS mergers are expected to have further mass ejection
by viscous heating
\citep{dessart09,fernandez13,fernandez15nsns,fernandez15bhns,shibata17}
that originates from magnetohydrodynamic turbulence
\citep{price06,kiuchi14,kiuchi15,giacomazzo15,ciolfi17,siegel17},
and subdominantly by neutrino heating \citep{wanajo12,perego14,fujibayashi17}
and nuclear recombination \citep{fernandez13}.
These components are as a whole denoted as ``post-merger'' ejecta in this paper.
The post-merger ejecta can consist of less neutron rich material
than in the dynamical ejecta \citep{just15,martin15,wu16,lippuner17};
neutrino absorption as well as a high temperature caused by viscous heating
makes ejected material less neutron rich
or electron fraction $\Ye$ (number of protons per nucleon) higher.
If the ejecta are free from Lanthanide elements,
the emission from post-merger ejecta can
be brighter and bluer,
which can be called ``blue kilonova'' \citep{metzger14,kasen15}.
However, due to the lack of atomic data of $r$-process elements,
previous studies assume opacities of Fe for Lanthanide-free ejecta.
To predict emission properties of kilonova,
systematic atomic data for $r$-process elements are important
\citep[see][]{kasen13,fontes17,wollaeger17}.

In this paper, we newly perform atomic structure calculations for
selected $r$-process elements.
Using these data, we perform radiative transfer simulations
and study the impact of element abundances to kilonova emission.
In Section \ref{sec:atomic}, we show methods and results of
our atomic structure calculations.
In Section \ref{sec:opacity}, we calculate opacities with these atomic
data and discuss the dependence on the elements.
We then apply our data for radiative transfer simulations
in Section \ref{sec:simulations}, and show light curves of kilonova
from dynamical and post-merger ejecta of NS mergers.
Finally we give summary in Section \ref{sec:summary}.

\begin{figure}
\begin{center}
  \includegraphics[scale=1.0]{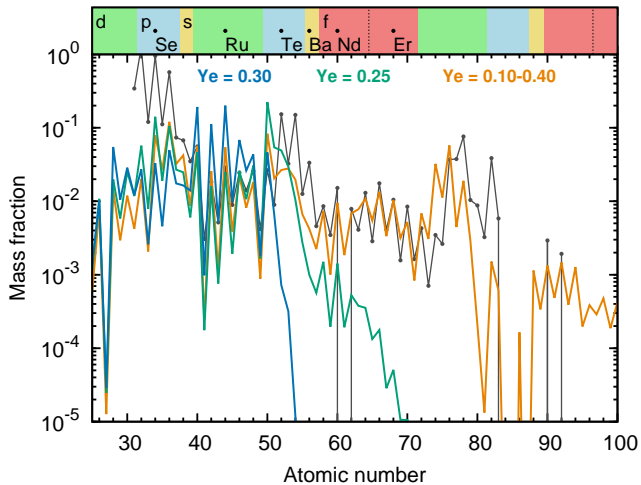}
\caption{
  Element abundances in the ejecta of NS mergers
  at $t=1$ day after the merger.
  The orange line shows abundances for dynamical ejecta \citep{wanajo14},
  which is derived by averaging the nucleosynthesis
  results of $\Ye = 0.10-0.40$ assuming a flat $\Ye$ distribution.
  The blue and green lines show the nucleosynthesis results
  from trajectories of $\Ye = 0.25$ and 0.30, respectively,
  which represent the abundance patterns of high-$\Ye$ post-merger ejecta.
  Black points connected with the line
  show the solar abundance ratios of $r$-process elements
  \citep{simmerer04}.
 \label{fig:abundance}}
\end{center}
\end{figure}

\section{Atomic Structure Calculations}
\label{sec:atomic}

We perform atomic structure calculations for
Se ($Z=34$), Ru ($Z=44$), Te ($Z=52$), Ba ($Z=56$), Nd ($Z=60$) and Er ($Z=68$).
These elements are selected to systematically study the
opacities of elements with different open shells:
Ba is an open s-shell element,
Se and Te are open p-shell elements, Ru is an open d-shell element,
and Nd and Er are open f-shell elements.
We focus on neutral atom and singly and doubly ionized ions 
because these ionization states are most common in kilonova
at $t \gsim 1$ day after the merger \citep{kasen13,tanaka13}.

In Figure \ref{fig:abundance},
these elements are shown with three different abundance patterns in
the ejecta of NS mergers.
While relativistic simulations of NS mergers predict wide ranges of $\Ye$
between 0.05 and 0.45, the detailed $\Ye$ distributions depend on the NS masses
and their ratios as well as the adopted nuclear equations of state
\citep{sekiguchi15,sekiguchi16}.
In this paper, we assume a flat mass distribution
between $\Ye =$ 0.10 and 0.40 as representative of dynamical ejecta.
As shown in Figure \ref{fig:abundance} (orange line),
the dynamical ejecta consist of a wide range of $r$-process elements from the first
($Z = 34$) to third ($Z = 78$) abundance peaks.
For the post-merger ejecta, we consider single $\Ye$ models of 0.25 (green)
and 0.30 (blue) for simplicity.
The former represents a case that contains the second ($Z = 52$)
abundance peak and a small amount of Lanthanides.
The latter is a Lanthanide-free model without elements of $Z > 50$.
For all the models, the nucleosynthesis abundances of each $\Ye$ are taken from
\citet{wanajo14}.

For the atomic structure calculations,
we use two different codes,
HULLAC \citep{bar-shalom01} and GRASP2K \citep{jonsson13}.
The HULLAC code, which employs a parametric potential method,
is used to provide atomic data for many elements
while the GRASP2K code, which enables more ab-initio calculations
based on the multiconfiguration Dirac-Hartree-Fock (MCDHF) method,
is used to provide benchmark calculations for a few elements.
Such benchmark calculations are important because
systematic improvement of the accuracies is not always obtained
with the HULLAC code especially when little data are available in NIST Atomic
Spectra Database (ASD, \citealt{kramida15}).
By using these two codes, we also study the influence of
the accuracies of atomic calculations to the opacities.
Tables \ref{tab:hullac} and \ref{tab:grasp} summarize
the list of ions for atomic structure calculations.
In the following sections, we describe our methods to calculate the
atomic structures and transition probabilities.

\begin{deluxetable*}{llrrrr} 
\tablewidth{0pt}
\tablecaption{Summary of HULLAC calculations}
\tablehead{
  Ion      & Configurations  & Number of levels  & Number of lines  & Subset 1$^a$ & Subset 2$^b$
}
\startdata
\ion{Se}{i}   &   ${\bf 4s^2 4p^4}$, $4s^2 4p^3 (4d,4f,5-8l)^c$, $4s 4p^5$, $4s 4p^4 (4d,4f)$,   &     3076   &  973,168    & 2,395      & 654 \\
               &   $4s^2 4p^2 (4d^2, 4d 4f,4f^2)$, $4s 4p^3 (4d^2,4d4f,4f^2)$        &    &    &    &  \\
\ion{Se}{ii}  &   ${\bf 4s^2 4p^3}$, $4s^2 4p^2 (4d,4f,5-8l)^c$, $4s 4p^4$, $4s 4p^3 (4d,4f)$,   &      2181   &   511,911      & 1,978      & 584 \\
               &   $4s^2 4p (4d^2, 4d 4f,4f^2)$, $4s 4p^2 (4d^2,4d4f,4f^2)$        &    &    &    &  \\
\ion{Se}{iii} &    ${\bf 4s^2 4p^2}$, $4s^2 4p (4d,4f,5-8l)^c$, $4s 4p^3$, $4s 4p^2 (4d,4f)$,     &     922    &   92,132      & 2,286      & 882 \\
               &   $4s^2 (4d^2, 4d 4f,4f^2)$, $4s 4p (4d^2,4d4f,4f^2)$        &    &    &    &  \\
\ion{Ru}{i}   &   ${\bf 4d^7 5s}$, ${\bf 4d^6 5s^6}$, ${\bf 4d^8}$, $4d^7 (5p,5d,6s,6p)$,         &    1,545  &  250,476   & 49,181   & 20,350 \\
                &   $4d^6 5s (5p,5d,6s)$                                   &                    &                     &                    & \\
\ion{Ru}{ii}  &   ${\bf 4d^7}$, $4d^6 (5s-5d,6s,6p)$               &     818   &   76,592   & 27,976   & 14,073 \\
\ion{Ru}{iii} &    ${\bf 4d^6}$, $4d^5 (5s-5d,6s)$                &    728    &   49,066   & 30,628   & 17,451 \\
\ion{Te}{i}   &   ${\bf 5s^2 5p^4}$, ${\bf 5s^2 5p^3 (4f,5d,5f,6s-6f,7s-7d,8s)}$,               &     329   &   14,482   & 410      & 348 \\
               &   $5s 5p^5$                      &              &              &              & \\
\ion{Te}{ii}  &  ${\bf 5s^2 5p^3}$, ${\bf 5s^2 5p^2 (4f,5d,5f,6s-6f,7s-7d,8s)}$,                &    253    &   9,167    & 705      & 569 \\
               &   $5s 5p^4$                      &              &              &              & \\
\ion{Te}{iii} &  ${\bf 5s^2 5p^2}$, ${\bf 5s^2 5p (5d,6s-6d,7s)}$,  $5s 5p^3$        &     57    &    419     & 249      & 227 \\
\ion{Nd}{i}   &  ${\bf 4f^4 6s^2}$, ${\bf 4f^4 6s(5d,6p,7s)}$, ${\bf 4f^4 5d^2}$, ${\bf 4f^4 5d 6p}$,    &  31,358   & 70,366,259 & 12,365,070 & 2,804,079 \\
               &   $4f^3 5d 6s^2$, $4f^3 5d^2 (6s,6p)$, $4f^3 5d 6s 6p$       &         &         &        &  \\
\ion{Nd}{ii}  &  ${\bf 4f^4 6s}$, ${\bf 4f^4 5d}$, $4f^4 6p$, $4f^3 6s (5d,6p)$,                 &  6,888    & 3,951,882 & 3,682,300  & 1,287,145 \\
               &   $4f^3 5d^2$, $4f^3 5d 6p$                    &        &        &        & \\
\ion{Nd}{iii} &  ${\bf 4f^4}$, $4f^3 (5d,6s,6p)$, $4f^2 5d^2$, $4f^2 5d (6s,6p)$,               &  2252     &  458,161   & 303,021    & 136,248 \\
               &  $4f^2 6s6p$        &       &          &       & \\
\ion{Er}{i}   &   ${\bf 4f^{12} 6s^2}$, ${\bf 4f^{12} 6s (5d,6p,6d,7s,8s)}$,                    &   10,535  &  9,247,777 & 443,566    & 129,713 \\
               &   $4f^{11} 6s^2 (5d,6p)$, $4f^{11} 5d^2 6s$, $4f^{11} 5d 6s (6p,7s)$                                                     &           &            &            & \\
\ion{Er}{ii}  &   ${\bf 4f^{12} 6s}$, $4f^{12} (5d,6p)$,  $4f^{11} 6s^2$, $4f^{11} 6s (5d,6p)$,             &   5,333   & 2,432,665  & 1,713,258   & 489,383  \\
              &    $4f^{11} 5d^2$, $4f^{11} 5d 6p$                            &            &           &          & \\
\ion{Er}{iii} &   ${\bf 4f^{12}}$, $4f^{11} (5d,6s,6p)$               &  723      &   42,671   & 41,843     & 16,787 \\ 
\enddata
\tablecomments{  
  Configurations taken into account for optimization of central-field potentials are indicated by bold letters (see text).
  $^a$ Number of lines whose lower level energy is $E_1 < 5$, 10, 15 eV for
  neutral atom and singly and doubly ionized ions, respectively.\\
  $^b$ Number of lines with $ \log (gf_l) \ge -3.0$ in Subset 1. \\
  $^c$ $(5 - 8l)$ stands for single orbitals in all nl shells with $n=5 - 8$ and$l= 0$ - $n-1$.
}
\label{tab:hullac}
\end{deluxetable*}

\begin{figure*}
\begin{center}
\begin{tabular}{cc}
\includegraphics[scale=0.9]{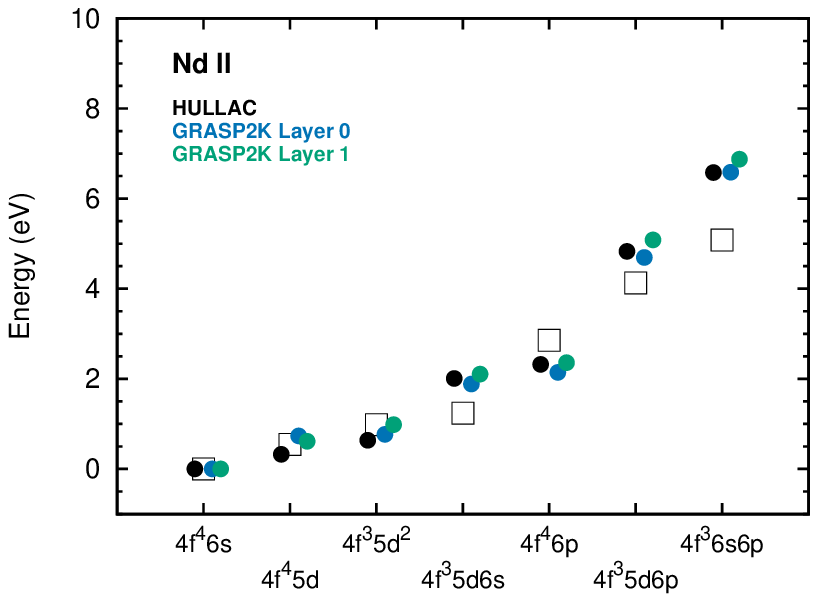} &
\includegraphics[scale=0.9]{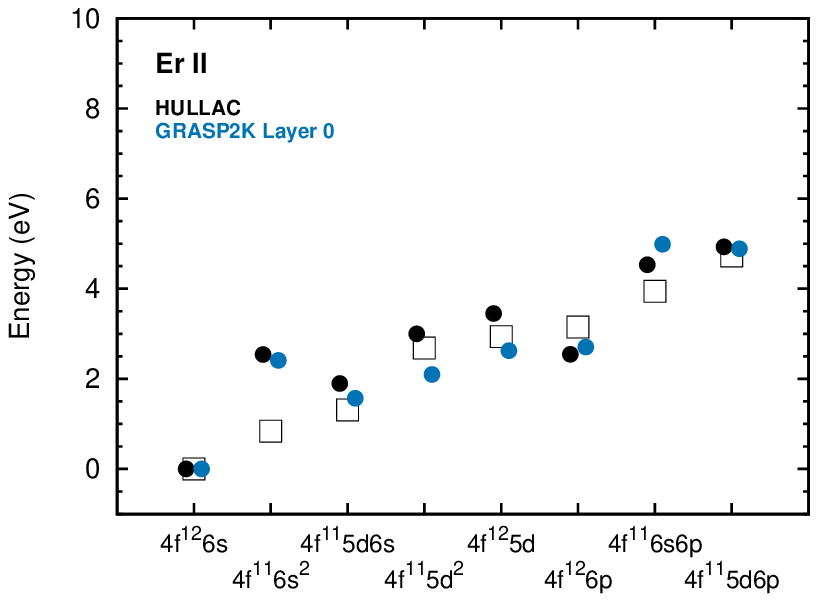} \\
\includegraphics[scale=0.9]{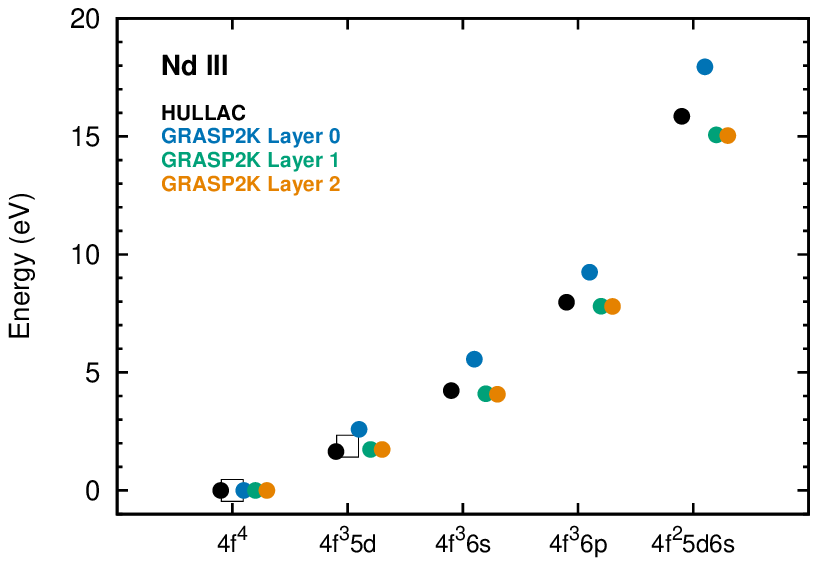} &
\includegraphics[scale=0.9]{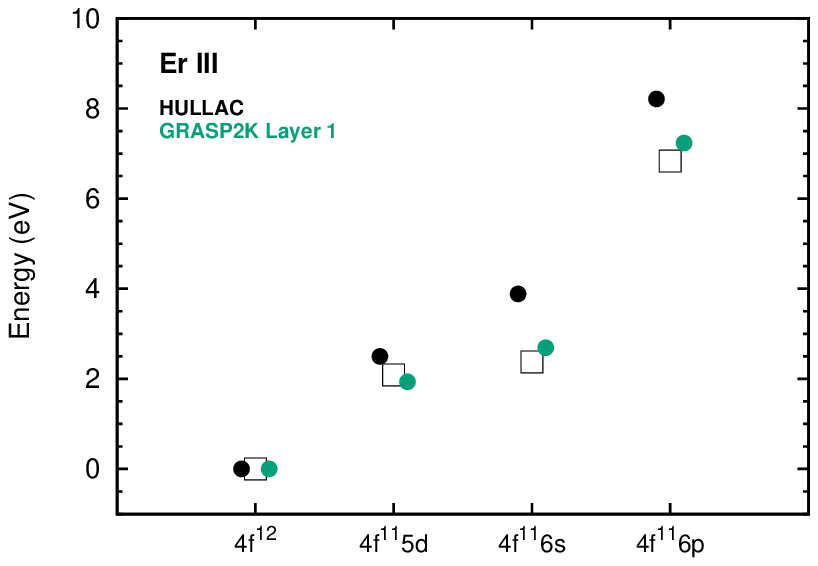}
\end{tabular}
\caption{
  The excitation energy of the lowest energy levels for
  each electron configuration.
  Black circles show HULLAC calculations while blue, green, and orange
  circles show GRASP2K calculations with different strategies.
  The data from the ASD \citep{kramida15} are shown in open squares for comparison.
\label{fig:Elevel}}
\end{center}
\end{figure*}

\subsection{HULLAC}

HULLAC (Hebrew University Lawrence Livermore Atomic Code, \citealt{bar-shalom01})
is an integrated code for calculating atomic structures
and cross sections for modeling of atomic processes in plasmas and emission spectra.
The latest version (9-601k) of HULLAC is used in the present work to provide atomic data
for Se {\sc i-iii}, Ru {\sc i-iii}, Te {\sc i-iii}, Nd {\sc i-iii}, and Er {\sc i-iii}.
In HULLAC, fully relativistic orbitals are used for calculations of atomic energy levels and radiative transition probabilities.
The orbital functions $\varphi_{nljm}$ are solutions of the single electron Dirac equation with a local central-field potential $U(r)$ which represents a nuclear field and a spherically averaged interaction with other electrons in atoms,
\begin{equation}
\left[  c {\bm{ \alpha }} \cdot
                    {\bm{ p }}
         + (\beta -1)c^2 + U(r) \right ] \varphi_{nljm} = \varepsilon_{nlj}\varphi_{nljm},
\end{equation}
where $\bm{\alpha}$ and $\beta$ are the $4 \times 4$ Dirac matrices, $c$ the speed of light in atomic units, and $\varepsilon_{nlj}$ an orbital energy
associated with each principal quantum number $n$,
azimuthal quantum number $l$, and angular quantum numbers $j$ and $m$.
$N$-electron configuration state functions (CSFs) are constructed by coupled anti-symmetrized products of the orbital functions using Racah algebra. 
The $jj$-coupling scheme is used in HULLAC.

The total Hamiltonian, which consists of the $N$-electron Dirac kinetic energy, the nuclear field potential, the inter-electronic interaction potential including the magnetic and retardation effects (the Breit interaction), and
quantum electrodynamic (QED) energy corrections, is diagonalized with multi-CSFs (relativistic configuration interaction: RCI).
Atomic energy levels are obtained from eigenvalues of the total Hamiltonian.
Atomic wave functions of the energy levels are expressed as linear combinations of CSFs.
Electric-dipole transition probabilities between two energy levels are obtained from the transition moments of the wave functions in length (Babushkin) gauge.
In Table \ref{tab:hullac}, the configuration set used in the diagonalization
as well as the number of energy levels and transitions of each ion are summarized.
The ground state configuration is indicated at the top.
It is noted that the configuration set in the present calculations
should be read as minimal.
Extended sets of configurations are used for Se to get improved energy levels. 

The optimal central-field potential is obtained such that energy levels of the ground state and a few excited states agree with those in the ASD.
The electron charge distribution function of $q$ electrons in an $nl$ shell is expressed by the density of the Slater-type orbital as,
\begin{equation}
\rho(r)=-4\pi r^2 q A \left [ r^{l+1}\exp{\left (-\alpha r/2 \right )} \right ]^2,
\end{equation}
where $A$ is a normalization factor and $\alpha$ values represent average radii of the Slater-type orbital. 
The central-field potential for this electron charge distribution and the nuclear charge distribution $Z \delta (r)$ seen by an external electron is obtained from the Poisson equation with the boundary condition, $U(r) |_{r \rightarrow \infty} = \left ( Z-q \right ) / r$.
Occupancy of each Slater-type orbital is naively chosen as the ground state configuration of the next higher charge state.
The ground state configuration for each ion is as given in the ASD.
Alternative occupancies will give different electron charge density distributions which result in different central-field potentials.
In some cases, such alternative occupancies are used to improve results.
For \ion{Ru}{i}, an alternative occupancy [Kr]~$4d^5 5s^2$ gives deeper and quasi-degenerate $4d$ and $5s$ orbital energies resulting in a better agreement with the energy levels of the ASD.
Similarly, alternative occupancies [Xe]~$4f^3 6s$ and [Cd]~$5p^5 4f^{12}$ are used for \ion{Nd}{ii} and \ion{Er}{iii}, respectively, in the present calculations.

The $\alpha$ values which minimize first-order configuration average energies of the ground state and low-lying excited states are chosen.
Such $\alpha$ values depend on excited state configurations added in the first-order energies to be minimized.
We choose the excited state configurations by single and double substitutions of valence and sub-valence orbitals from the ground state configuration.
The ground state for each ion as well as 
the excited state configurations taken into account for the energy
minimization are indicated by bold letters in Table \ref{tab:hullac}.
Getting correct energy levels by this semi-empirical optimization takes a less computational time with limited computational resources, although systematic improvement of the results without a benchmark is not always possible.
Results of a few lowest excited energy levels deviate from those of the
ASD about 10\% at most for Se and Te.
However, we cannot obtain such close agreements for Ru reflecting
complexity of the atomic structures with open d shells.
Results of the energies for Nd {\sc ii-iii} and Er {\sc ii-iii} are shown 
in Figure \ref{fig:Elevel} and discussed in the following section.

\begin{deluxetable*}{c c c c c c c c c c c }
\tablewidth{0pt}
\tablecaption{Summary of GRASP2K calculations}
\tablehead{
Ion & Inact. & Ground & \multicolumn{2}{c}{multi-reference set} & Active set &  \multicolumn{2}{c}{Number of levels} &&  \multicolumn{2}{c}{N$_{CSFs}$} \\ 
\cline{4-5} \cline{7-8} \cline{10-11}
&core&conf.& Even & Odd & & Even & Odd && Even & Odd 
}
\startdata
\ion{Ba}{ii}& [Kr]& $4d^{10} 5s^2 5p^6 6s$ & $ 4d^{10}5s^2 5p^6 ns$ & $4d^{10} 5s^2 5p^6 np$ &$\{ns,np,(n-1)d,$ & 75 & 60 && 838,672 & 614,880\\
    &     &                 & $4d^{10} 5s^2 5p^6 (n-1)d $ & $4d^{10} 5s^2 5p^6 (n-2)f$& $(n-2)f,(n-1)g\}$ \\
    &     &                 & $4d^{10} 5s^2 5p^6 (n-1)g $\\
    &     &                 & $n=6-20$\\
\ion{Ba}{iii}&[Kr]& $4d^{10} 5s^2 5p^6$ & $4d^{10} 5s^2 5p^6$ $^a$ & $4d^{10} 5s^2 5p^5 ns$ & $\{ns,np,(n-1)d,$ & 409 & 504 && 70,067& 71,388\\
     &    &                 & $4d^{10} 5s^2 5p^5 np$ & $4d^{10} 5s^2 5p^5 (n-1)d$   & $(n-2)f,(n-1)g$ \\
     &    &                 & $4d^{10} 5s^2 5p^5 (n-2)f$ & $4d^{10} 5s^2 5p^5 (n-1)g$ & $nh ^a$\}  \\
     &    &                 & $4d^{10} 5s^2 5p^5 nh$ $^a$ &                           &           \\
     &    &                 & $n=6-23$\\
\ion{Nd}{ii} & [Xe] & $4f^4 6s$ & $4f^4 6s$, $4f^4 5d$   & $4f^3 5d^2$, $4f^4 6p$ & \{7s,7p,6d,5f\} & 3,890 & 2,998 && 24,568 & 23,966 \\
&&& $4f^3 5d 6p$, $4f^3 6s 6p$ & $4f^3 5d 6s$ \\ \\[-0.3cm]

\ion{Nd}{iii} & [Xe] & $4f^4$ & $4f^4$, $4f^3 6p$   & $4f^3 5d$, $4f^3 6s$ & \{8s,8p,7d,6g,6h\} & 1,020 & 468 && 173,816 & 114,621 \\
&&& $4f^2 5d^2$, $4f^2 5d 6s$ $^b$ \\ \\[-0.3cm]

&&&  \\ \\[-0.3cm]

\ion{Er}{ii} & [Xe] & $4f^{12} 6s$ & $4f^{12} 6s$, $4f^{12} 5d$   & $4f^{11} 5d 6s$, $4f^{11} 5d^2$ & \{7s,7p,6d,5f\} & 2,836 & 2,497 && 22,460 & 21,731 \\
&&&  $4f^{11} 5d 6p$, $4f^{11} 6s 6p$ & $4f^{11} 6s^2$, $4f^{12} 6p$ \\ \\[-0.3cm]

\ion{Er}{iii} & [Xe] & $4f^{12}$ & $4f^{12}$, $4f^{11} 6p$   & $4f^{11} 5d$, $4f^{11} 6s$  & \{7s,7p,6d,5f\} & 255 & 468 && 503,824 & 842,643 \\ \\[-0.3cm]

\enddata
\tablecomments{
  Summary of the MCDHF and RCI calculations indicating inactive core, multi-reference set, active set, number of calculated levels and number of configuration state functions (N$_{CSFs}$) in the final list for each of parity. \\
  $^a$ Ground state and states of the configuration $4d^{10} 5s^2 5p^5 nh$ as the $nh$
orbital were excluded from the computations for $n=7-23$.\\
  $^b$ To match with the configurations used by \citet{kasen13},
  the $4f^2 5d 6p$ configuration is not included in the GRASP2K calculations
  while it is included in the HULLAC calculations.
  As a result, the HULLAC results have more energy levels (2252) compared
  with those of GRASP2K (1488).
  This difference does not have a big impact on the opacities
  as the energy levels from the configuration is rather high.
} 
\label{tab:grasp} 
\end{deluxetable*}

\subsection{GRASP2K}

GRASP2K \citep{jonsson13} is used to provide atomic data for 
Ba {\sc ii-iii}, Nd {\sc ii-iii}, and Er {\sc ii-iii}.
GRASP2K is based on the MCDHF and RCI methods taking into account Breit and QED
corrections \citep{grant07,froesefischer16}.
Based on the Dirac-Coulomb Hamiltonian
\begin{equation}
H_{DC} = \sum_{i=1}^N \left( c {\bm{ \alpha }}_i \cdot
                    {\bm{ p }}_i
         + (\beta_i -1)c^2 + V^N_i \right)
         + \sum_{i>j}^N \frac{1}{r_{ij}},
\end{equation}
where $V^N$ is the monopole part of the electron-nucleus Coulomb interaction.
The atomic state functions (ASFs) are obtained as linear
combinations of symmetry adapted CSFs.
The  CSFs are built from products of one-electron Dirac orbitals. 
Based on a weighted energy average of several states,
the so called extended optimal level (EOL) scheme \citep{dyall89},
both the radial parts of the Dirac orbitals and the 
expansion coefficients are optimized self-consistently in the
relativistic self-consistent field procedure.
In the present calculations, ASFs are obtained as
expansions over $jj$-coupled CSFs.
To provide the $LSJ$ labeling system
the ASFs are transformed from a $jj$-coupled CSF basis into
an $LSJ$-coupled CSF basis using the
method provided by \citet{gaigalas17}.

The MCDHF calculations are followed by
RCI calculations,
including the Breit interaction and leading QED effects.
Note that, for \ion{Nd}{ii} and \ion{Er}{ii},
only MCDHF calculations are performed.
Radiative transition data
(transition probabilities, oscillator strengths) between two states
built on different and independently
optimized orbital sets are calculated by means of the biorthonormal
transformation method \citep{olsen95}.
For electric dipole and quadrupole (E1 and E2) transitions,
we use the Babushkin gauge as in the HULLAC calculations.

In the Table \ref{tab:grasp}, we give a summary of
the MCDHF and RCI calculations for each of ion.
As a starting point, MCDHF calculations are performed in the EOL
scheme for the states of the ground configuration.
The wave functions from these calculations are taken as the initial one to calculate
even and odd states of multi-reference configurations.
The set of orbitals belonging to these multi-reference configurations
are referred to Layer 0.
After that the even and odd states are calculated separately.
Unless stated otherwise, in the present calculations
the inactive core for each of ions is mentioned in Table \ref{tab:grasp}.
The CSF expansions for states of each parity are
obtained by allowing single and
double substitutions from the multi-reference configurations up to
active orbital sets (see Table \ref{tab:grasp}). 
The configuration space was increased step by step with increasing layer number. 
The orbitals of previous layers 
are held fixed and only the orbitals of the new layer are allowed to vary.

\subsection{Results}

Figure \ref{fig:Elevel} shows the derived lowest energy
for each electron configuration
of \ion{Nd}{ii}, \ion{Nd}{iii}, \ion{Er}{ii}, and \ion{Er}{iii} ions.
For \ion{Nd}{ii}, both HULLAC and GRASP2K calculations show 
reasonable agreement with the data in the ASD (open squares).
Our calculations provides the correct orders of energy levels,
and the deviation from the energy in the ASD is about $\lsim 30 \%$
except for the ${\rm 4f^36s6p}$ configuration.
Overall agreement is slightly better than that obtained by \citet{kasen13}
with Autostructure code \citep{badnell11}.
For \ion{Nd}{iii}, the energy of the excited level is available only for
${\rm 4f^35d}$ and our calculations give an excellent agreement
except for the Layer 0 calculation of GRASP2K (blue points).

For the case of Er,
the agreement with the ASD is not as good as in the Nd ions,
which reflects the complexity of the Er ions.
For \ion{Er}{ii}, both HULLAC and GRASP2K calculations do not give
the correct order of energy:
the derived energy of ${\rm 4f^{11}6s^2}$ is too high.
On the other hand, for \ion{Er}{iii},
the orders of energy are reproduced well
although the number of available data in the ASD is small.
For the case of \ion{Er}{iii}, GRASP2K results give
a better agreement than HULLAC.
In the next section, we discuss the influence of these results to the opacities.

\begin{figure*}[t]
\begin{center}
\begin{tabular}{cc}
\includegraphics[scale=0.9]{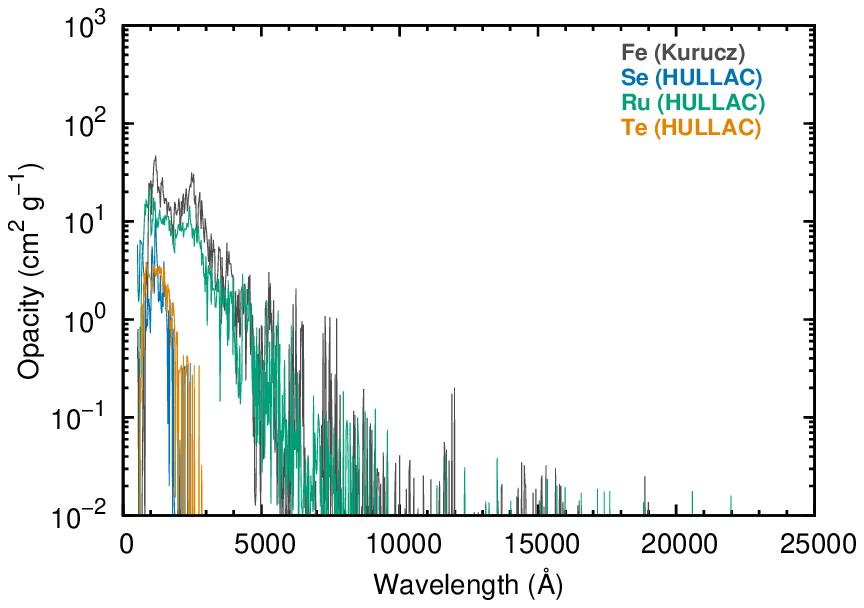} &
\includegraphics[scale=0.9]{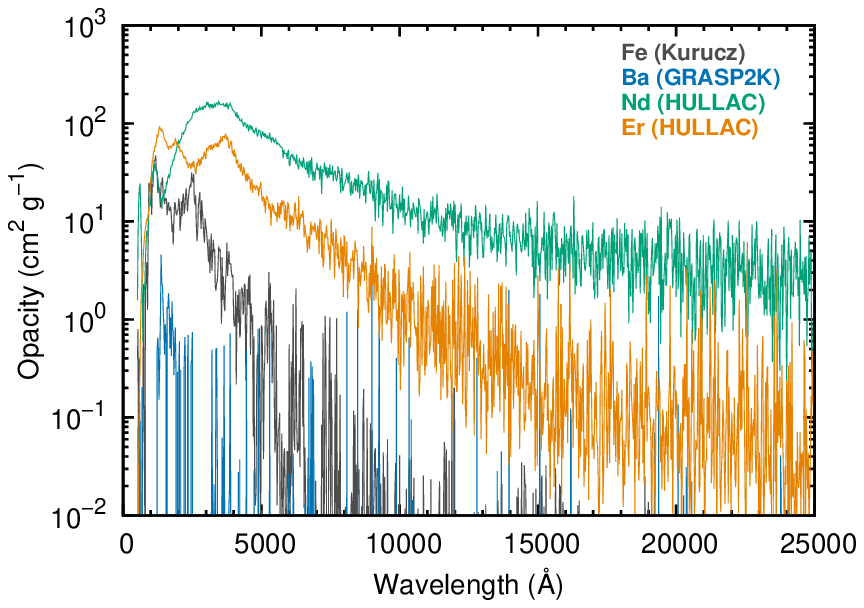} 
\end{tabular}
\caption{
  Line expansion opacities of Se, Ru, Te (left), Ba, Nd, and Er (right).
  The calculations assume $\rho = 1 \times 10^{-13} \ {\rm g \ cm^{-3}}$,
  $T = 5,000$ K, and $t=$ 1 day after the merger.
  The results are compared with the line expansion opacities of Fe
  calculated with Kurucz's line list.
\label{fig:opacity_T5}}
\end{center}
\end{figure*}

\begin{figure*}[t]
\begin{center}
\begin{tabular}{cc}
\includegraphics[scale=0.9]{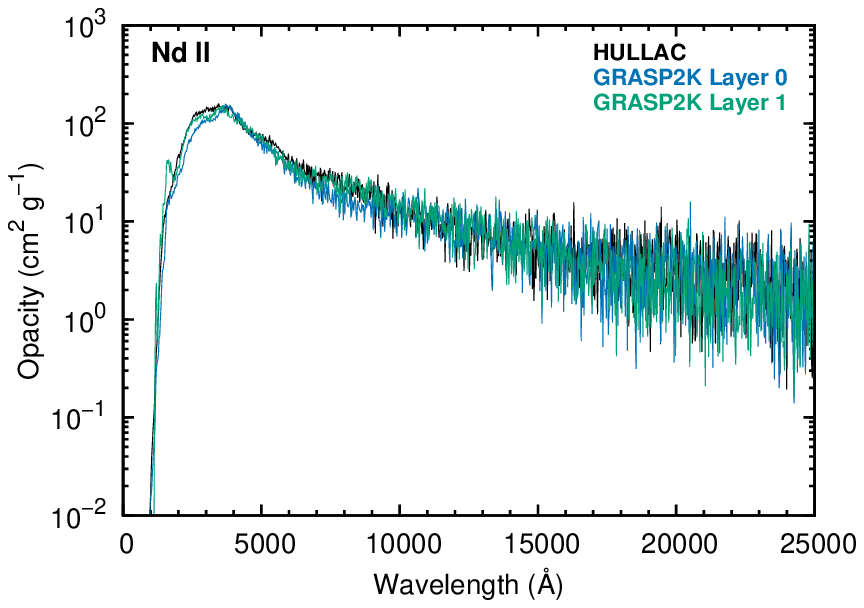} &
\includegraphics[scale=0.9]{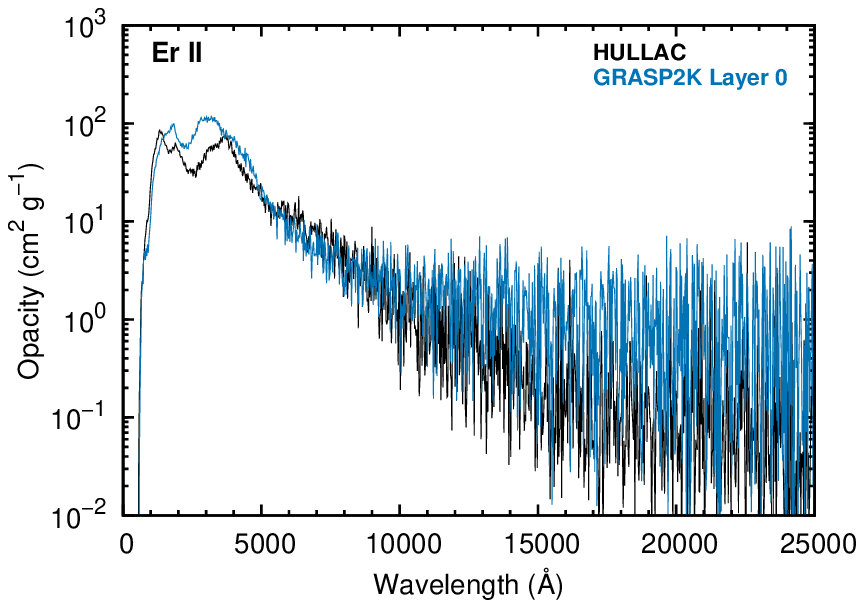} \\
\includegraphics[scale=0.9]{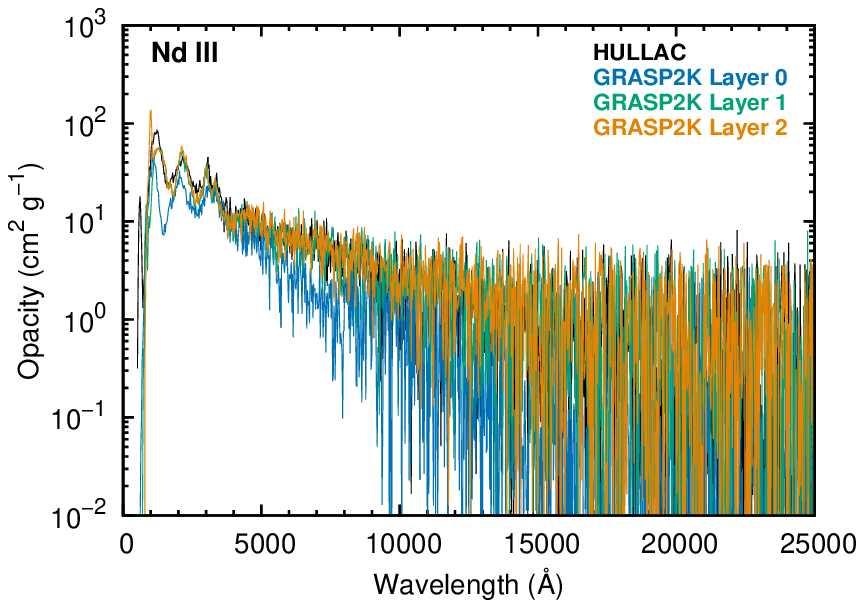} &
\includegraphics[scale=0.9]{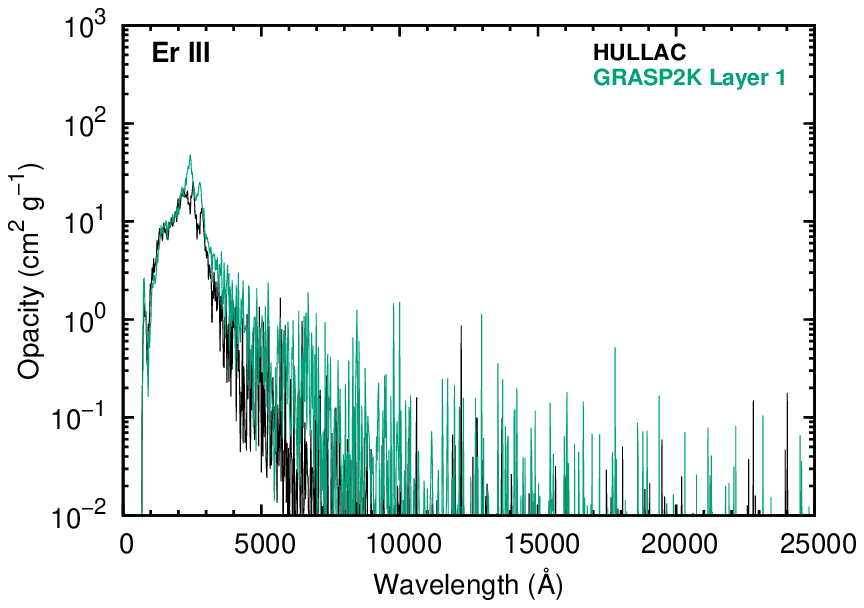}
\end{tabular}
\caption{
  Comparison of line expansion opacities
  between HULLAC and GRASP2K calculations.  
  For singly ionized ions (\ion{Nd}{ii} and \ion{Er}{ii}),
  the calculations assume $\rho = 1 \times 10^{-13} \ {\rm g \ cm^{-3}}$,
  $T = 5,000$ K, and $t=$ 1 day after the merger.
  For doubly ionized ions (\ion{Nd}{iii} and \ion{Er}{iii}),
  the calculations assume the same density at the same epoch but
  $T = 10,000$ K.
\label{fig:opacity_comp}}
\end{center}
\end{figure*}

\begin{figure}
\begin{center}
  \includegraphics[scale=0.9]{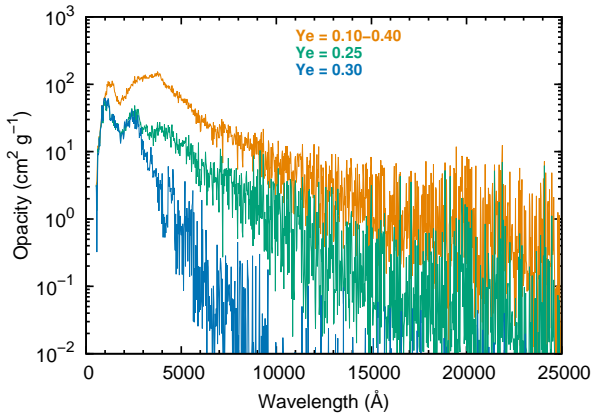} 
\caption{
  Line expansion opacities for mixture of elements
  in the ejecta of NS mergers (see Figure \ref{fig:abundance}).
  The orange line represents the opacity in the dynamical ejecta,
  which is calculated with the abundance pattern of $\Ye = 0.10-0.40$.
  The blue and green lines represent the opacities in the
  high-$\Ye$ post-merger ejecta,
  which are calculated with the abundance patterns of
  $\Ye = 0.25$ and 0.30, respectively.
  All the calculations assume $\rho = 1 \times 10^{-13} \ {\rm g \ cm^{-3}}$,
  $T = 5,000$ K, and $t=$ 1 day after the merger.
\label{fig:opacity_model}}
\end{center}
\end{figure}

\section{Opacity}
\label{sec:opacity}

We calculate the bound-bound opacities by using the
energy levels and the transition probabilities obtained from our atomic
structure calculations.
For bound-bound opacities, we use the formalism of expansion opacity
\citet{karp77,eastman93,kasen06};
\begin{equation}
\alpha_{\rm exp}^{\rm bb} (\lambda) = \frac{1}{ct} 
\sum_l \frac{\lambda_l}{\Delta \lambda} (1 - e^{- \tau_l}),
\end{equation}
which are also adopted by previous studies of kilonova simulations
\citep{kasen13,barnes13,tanaka13,tanaka14}.
Here $\tau_l$ is the Sobolev optical depth for a transition,
and it can be written as
\begin{equation}
  \tau_l = \frac{\pi e^2}{m_e c} f_l n_{l} t \lambda_l 
\end{equation}
for homologously expanding ejecta.
Here, $\lambda_l$ and $f_l$ are the wavelength and
the oscillator strength of the transition, respectively,
and $n_l$ is the number density of the lower level of the transition.
The summation in the expansion opacity
is taken for all the transition within a wavelength bin ($\Delta \lambda$).
The number density of each ion is calculated under the assumption
of local thermodynamic equilibrium,
and the population of the excited levels is calculated 
by assuming the Boltzmann distribution.

We confirm that there is an overall trend that
the bound-bound opacities of open f-shell (Lanthanide) elements
are higher than those of open s-shell, p-shell, and d-shell elements
over the wide wavelength range.
The opacities of open d-shell elements (Fe and Ru) concentrates on
the ultraviolet and optical wavelengths,
and those of open s-shell (Ba) and p-shell (Se and Te) elements also have a
similar trend with even lower opacities.
Figure \ref{fig:opacity_T5} shows the opacity of each element
calculated with $\rho = 1 \times 10^{-13} \ {\rm g \ cm^{-3}}$,
$T = 5,000$ K, and $t=$ 1 day after the merger.
For comparison, Fe opacities with Kurucz's line list
\citep{kurucz95} are also shown.
The opacities of Nd and Er (open f-shell) are much higher than
that of Fe (open d-shell).
The opacity of Ru (open d-shell) is similar to that of Fe,
which demonstrates the similarity in the opacity
for the elements with the same open shell.
The same is true for open p-shell;
the opacities of Se and Te are found to be similar.

The opacities from the two atomic codes agree reasonably well.
Figure \ref{fig:opacity_comp} shows
the line expansion opacities of
\ion{Nd}{ii}, \ion{Nd}{iii}, \ion{Er}{ii}, and \ion{Er}{iii}.
As expected from the good agreement in the energy levels
(Figure \ref{fig:Elevel}),
the opacities from HULLAC and GRASP2K are almost indistinguishable for
\ion{Nd}{ii} and \ion{Nd}{iii}.
It is noted that, only the Layer 0 calculations with GRASP2K
gives \ion{Nd}{iii} opacities lower than more realistic
(Layer 1 and Layer 2) calculations.
This is because the Layer 0 calculations give higher energy levels
(Figure \ref{fig:Elevel}),
which reduces the contribution of bound-bound transitions
involving excited levels for a given temperature.

For the Er ions, we find that two atomic codes give
larger discrepancies in the energy levels compared with the cases of Nd ions.
As in the case of \ion{Nd}{iii},
opacities of \ion{Er}{ii} and \ion{Er}{iii}
from HULLAC calculations are slightly smaller that those
from GRASP2K calculations 
because HULLAC calculations give slightly higher energy levels
for the excited energy levels.
However, the difference in the opacity is only up to a factor of about 2.
Therefore, we conclude that a relatively simplified calculations
with the HULLAC code
gives opacities with sufficient accuracies for astronomical applications.

Finally we calculate the opacities for mixture of elements.
We use the HULLAC results which cover more elements and ionization states.
Because we have atomic structure calculations for a small number of elements,
we assume the same bound-bound transition properties for 
the elements with the same open shell (see Figure \ref{fig:abundance}).
For open f-shell elements, the former and latter halfs
are replaced with Nd and Er, respectively.
For the heavy elements with $Z>71$,
we repeat to use the data of Ru, Te, Nd, and Er.
For the elements with $Z<32$, we use Kurucz's line list \citep{kurucz95}.
We neglect the contribution of open s-shell elements because
the total fraction of these elements are small in the ejecta (Figure \ref{fig:abundance})
and the opacities are subdominant (Figure \ref{fig:opacity_T5}).

As a result of high opacity of Lanthanide elements,
the opacities for the mixture of elements depends significantly on $\Ye$.
Figure \ref{fig:opacity_model} shows the line expansion opacity
for the element mixture in the dynamical ejecta ($\Ye = 0.10-0.40$)
and high-$\Ye$ ejecta ($\Ye = 0.25$ and $0.30$).
If the ejecta is completely Lanthanide free as in the case of $\Ye = 0.30$,
the line expansion opacity is smaller than that in the Lanthanide-rich ejecta
by a factor of $> 10$ near the middle of optical range ($\sim 5000$ \AA).
However, small inclusion of Lanthanide elements dramatically enhances
the opacities as shown in the case of $\Ye = 0.25$.
This demonstrates the importance of accurate $\Ye$ determination
in the merger simulations for the accurate prediction of kilonova signals.

\begin{figure}
\begin{center}
  \includegraphics[scale=1.0]{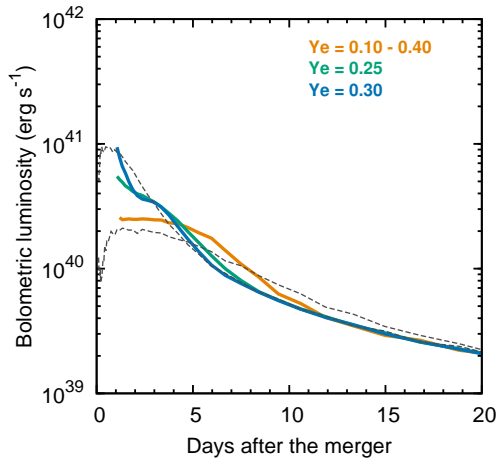} 
\caption{
Bolometric light curves of simple NS merger models
with $\Mej = 0.01 \Msun$ and $v_{\rm ch} = 0.1c$.
The solid curves show the results
with the wavelength-dependent radiative transfer
for different abundance ratios according to Figure \ref{fig:abundance}:
$\Ye = 0.10-0.40$ (orange), $\Ye = 0.25$ (green), and
$\Ye = 0.30$ (blue).
The dashed lines shows the results with the gray radiative transfer
with the gray opacity of 1.0 and 10.0 $\rm cm^2 \ g^{-1}$
from top to bottom.
For all the models, the analytic heating rates are used
and constant thermalization efficiency ($\epsilon = 0.25$) is assumed.
\label{fig:Lbol_NSM}}
\end{center}
\end{figure}

\section{Radiative Transfer Simulations}
\label{sec:simulations}

We perform radiative transfer simulations by using our new atomic data.
We use three-dimensional, time-dependent, wavelength-dependent
Monte Carlo radiative transfer code \citep{tanaka13}.
The code takes into account electron scattering
and bound-bound, bound-free, and free-free transitions as sources of opacity.
In the previous version of the code \citep{tanaka13,tanaka14,tanaka16},
we use the VALD database \citep{piskunov95,ryabchikova97,kupka99,kupka00}
for the bound-bound transitions,
while in this paper we use our atomic data presented
in Section \ref{sec:atomic} 
and treat element mixture by using representative elements
as described in Section \ref{sec:opacity}.
To save the memory space in the computation,
we use a subset of the line list including the transitions
whose lower level energy is $E_1 < 5$, 10, 15 eV for
neutral atom and singly and doubly ionized ions, respectively
(Subset 1 in Table \ref{tab:hullac})
and whose oscillator strengths are $ \log (gf_l) \ge -3.0$
(Subset 2 in Table \ref{tab:hullac}).
We confirm that the use of this subset does not significantly
affect the calculated light curves and spectra.

\subsection{Simple models}
\label{sec:simple}

To study the effect of the element abundances (or $\Ye$) on the light curves,
we calculate the light curves for the three different abundance patterns
displayed in Figure \ref{fig:abundance}.
For ease to extract the effect of opacities on the element abundances,
we employ a simple model of NS merger ejecta,
of which parameters are set to be the same for three cases.
The ejecta mass is taken to be $\Mej = 0.01 \Msun$.
The density structure of the ejecta is assumed to be spherical
with a power-law radial profile of
$\rho \propto r^{-3}$ from $v=0.05 c$ to $0.2c$
\citep{metzger10,tanaka13,metzger17}.
This model has a characteristic velocity of 
$v_{\rm ch} = \sqrt{2\KE/\Mej} = 0.1c$,
where $\KE$ is the kinetic energy of the ejecta.

For the heating rate of radioactive decays,
we adopt $2 \times 10^{10} t_{\rm d}^{-1.3}$ $\rm erg \ s^{-1}\ g^{-1}$
($t_{\rm d}$ is the time after the merger in days),
which gives a reasonable agreement with nucleosynthesis calculations
for a wide range of $\Ye$ \citep{wanajo14}.
We assume that a thermalization factor ($\epsilon$),
that is, a fraction of the decay energy deposited to the ejecta, 
is time-independent, and adopt $\epsilon = 0.25$, 
which is a typical value at a few days after the merger
\citep{barnes16,rosswog17}.

Figure \ref{fig:Lbol_NSM} shows the bolometric light curves of
the simple models.
Overall properties of the bolometric light curves can be understood
by the dependence on the opacities;
the characteristic timescale becomes shorter and
the luminosity becomes higher for smaller opacities.
As a result, the models with $\Ye =$ 0.25 and 0.30 has higher
bolometric luminosities at the first few days.

For comparison, we also show the results with gray transfer simulations
with $\kappa = 1.0$ and 10 $\ {\rm cm^2 \ g^{-1}}$.
It should be noted, however, that the gray opacities
give a reasonable approximation only for the bolometric luminosity.
Properties of multi-color light curves cannot be well described
by the gray opacities
as expected from the strong wavelength dependence of the opacities
(Figure \ref{fig:opacity_model}).

\begin{figure}
\begin{center}
\includegraphics[scale=1.0]{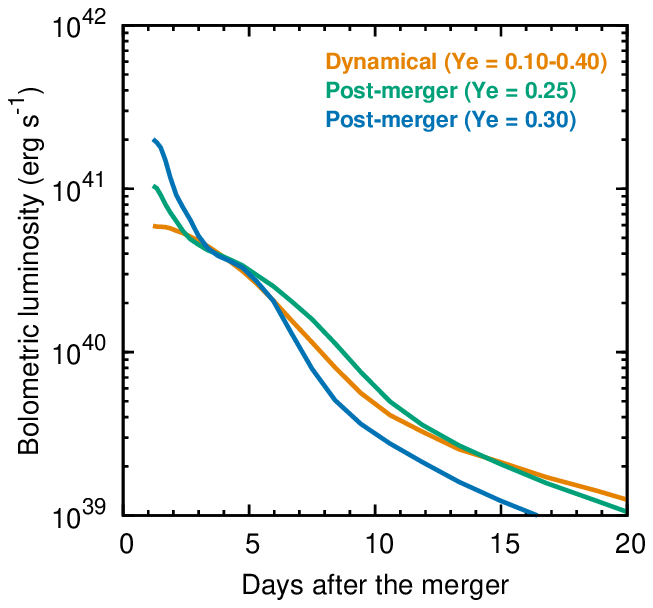}  
\caption{
Bolometric light curves of dynamical and post-merger ejecta models.
The orange line shows the NS merger model APR4-1215 \citep{hotokezaka13}
with $\Mej = 0.01 \Msun$ and the element abundances of
$\Ye = 0.10-0.40$ in Figure \ref{fig:abundance}.
Blue and green lines show the post-merger ejecta models 
(power-law density profile with $\Mej = 0.01 \Msun$
and $v_{\rm ch} = 0.05c$) with the element abundances
of $\Ye = 0.30$ and 0.25, respectively.
For all the models,
the heating rates from nucleosynthesis calculations \citep{wanajo14}
are used and the thermalization efficiencies \citep{barnes16}
are taken into account.
\label{fig:Lbol_model}}
\end{center}
\end{figure}

\begin{figure*}
\begin{center}
\begin{tabular}{cc}
\includegraphics[scale=1.0]{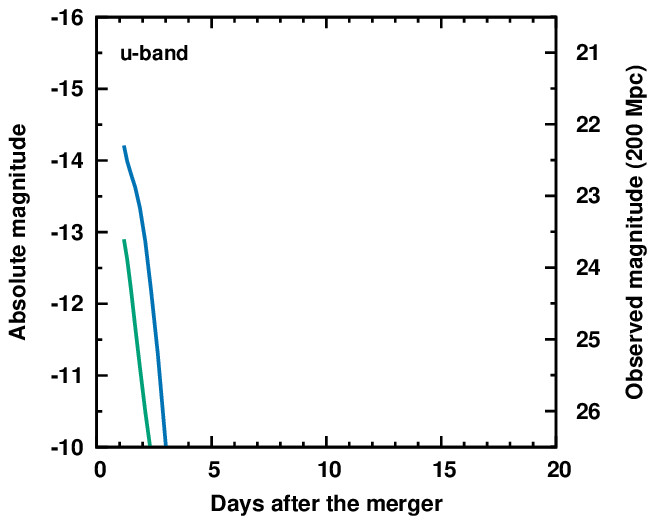} &
\includegraphics[scale=1.0]{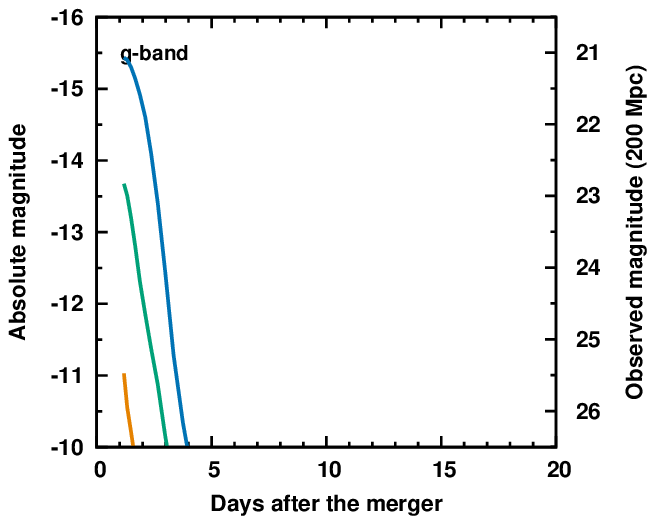} \\
\includegraphics[scale=1.0]{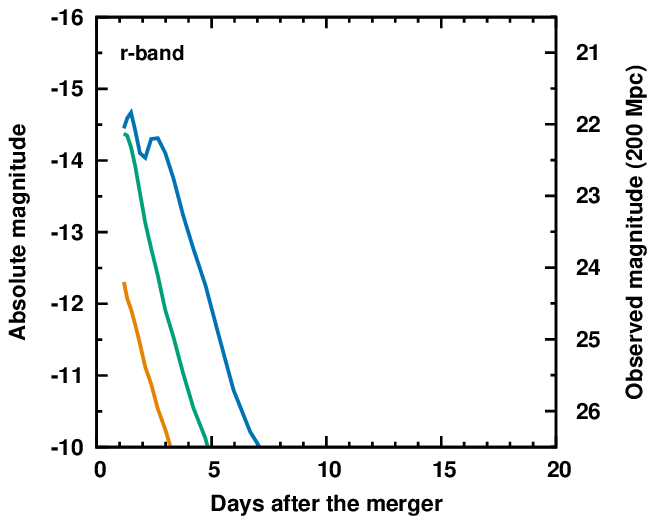} &
\includegraphics[scale=1.0]{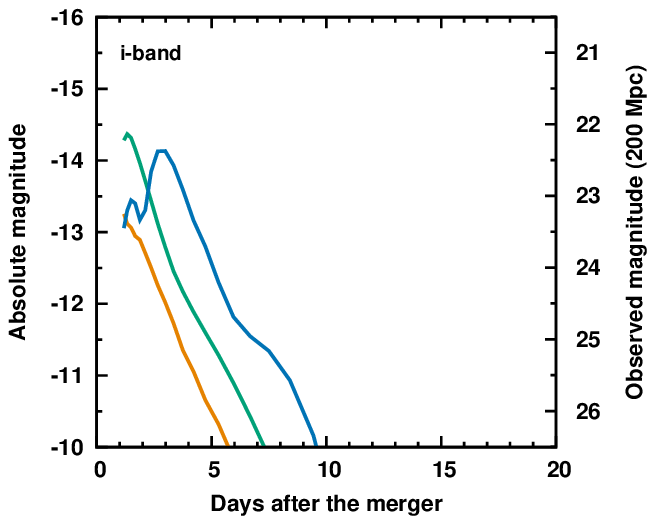} \\
\includegraphics[scale=1.0]{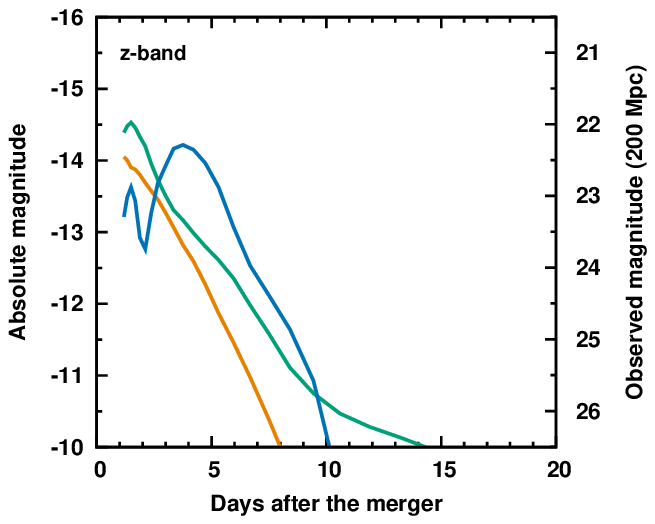} &
\includegraphics[scale=1.0]{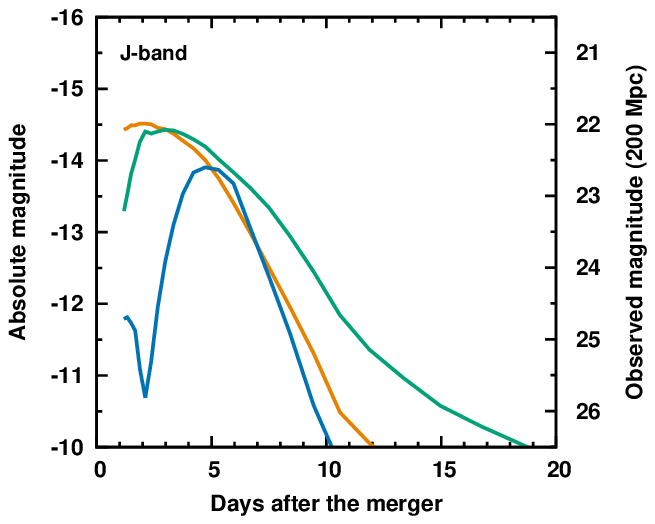} \\
\includegraphics[scale=1.0]{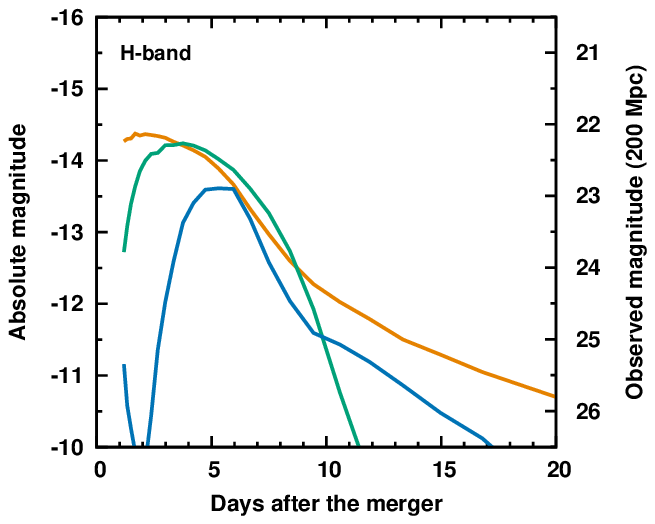} &
\includegraphics[scale=1.0]{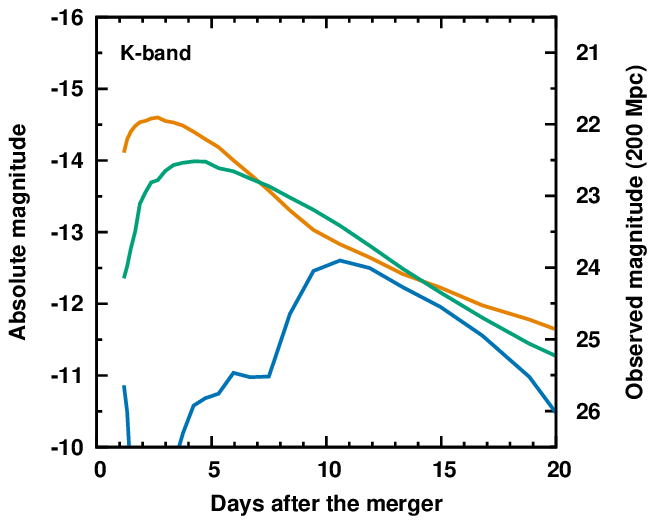} \\  
\end{tabular}
\caption{
Multi-color ($ugrizJHK$-band) light curves 
for the dynamical ejecta model APR4-1215 with $\Ye = 0.10-0.40$ (orange)
and the post-merger ejecta models with $\Ye = 0.30$ (blue) and 0.25 (green).
The vertical axis on the left shows the absolute magnitude
while that on the right the observed magnitude at 200 Mpc.
All the magnitudes are given in AB magnitudes.
\label{fig:mag}}
\end{center}
\end{figure*}

\begin{figure}
\begin{center}
  \includegraphics[scale=1.2]{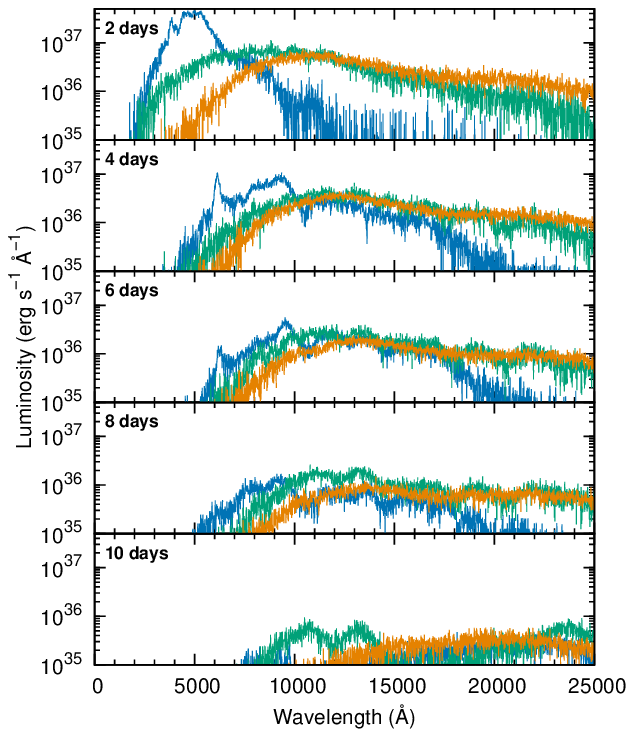}
\caption{
  Spectra of dynamical and post-merger ejecta models at
  $t = 2$, 4, 6, 8, and 10 days after the merger.
The orange line shows the NS merger model APR4-1215 \citep{hotokezaka13}
with $\Mej = 0.01 \Msun$ and the element abundances of
$\Ye = 0.10-0.40$ in Figure \ref{fig:abundance}.
Blue and green lines show the post-merger ejecta models 
(power-law density profile with $\Mej = 0.01 \Msun$
and $v_{\rm ch} = 0.05c$) with the element abundances
of $\Ye = 0.30$ and 0.25, respectively.
\label{fig:spec}}
\end{center}
\end{figure}

\subsection{Realistic Models}
\label{sec:realistic}

We perform radiative transfer simulations for
more realistic models of dynamical and high-$\Ye$ post-merger ejecta.
For the dynamical ejecta model,
we use the density structure from the results of numerical relativity
simulations by \citet{hotokezaka13}.
As a representative case, we use the APR4-1215 model,
which is the merger of two NSs with the gravitational masses of
$1.2 \Msun$ and $1.5 \Msun$.
The ejecta mass is $\Mej = 0.01 \Msun$
and the characteristic velocity is $v_{\rm ch} = 0.24c$.

The dynamical ejecta have a wide range of $\Ye$,
which composes of the low-$\Ye$ tidally disrupted component and
the high-$\Ye$ component by shock heating and/or neutrino absorption
\citep{wanajo14,sekiguchi15,goriely15,sekiguchi16,radice16,foucart16,lehner16}.
Therefore, we approximate the calculated abundance patterns
in \citet{wanajo14} by a flat $\Ye$ distribution from 0.10 to 0.40
as described in Section \ref{sec:atomic}
(orange line in Figure \ref{fig:abundance}).
We assume spatially homogeneous element distribution for simplicity
although merger simulations suggest that the polar regions
consist mainly of high $\Ye$ material \citep[\eg][]{sekiguchi16}.

For models of post-merger ejecta,
which originate from several possible mechanisms such as
viscosity, neutrino heating, and nuclear recombination,
we keep using a simple spherical model with a power-law profile
as in Section \ref{sec:simple},
instead of using results of numerical simulations.
We set $\Mej = 0.01 \Msun$ as a representative case.
Because the typical velocity of the post-merger ejecta
tends to be lower than that of the dynamical ejecta,
we set the velocity range from $v = 0.02c$ to $0.1c$,
which gives a characteristic velocity of $v_{\rm ch} = 0.05c$.
The element abundances in the ejecta can vary
depending on the detailed mechanisms of mass ejection.
In this paper, to study the effect of Lanthanide-free ejecta,
we assume relatively high $\Ye$,
that is, $\Ye =$ 0.25 and 0.30 as shown in Figure \ref{fig:abundance}.
The spatial distribution of the elements are assumed to be homogeneous
as in the dynamical ejecta.

For both dynamical and post-merger ejecta models,
we use the heating rates from nucleosynthesis calculations
for relevant $\Ye$ in \citet{wanajo14}.
The radioactive decay and subsequent energy deposition
from $\beta$-decay, $\alpha$-decay, and fission are taken into account.
For the $\beta$-decay, 45\% and 20\% of energy are assumed to be
carried out by $\gamma$-rays and $\beta$ particles, respectively.
Then, thermalization efficiencies of
$\gamma$-rays, $\alpha$ particles, $\beta$ particles,
and fission fragments are independently evaluated
by using analytic descriptions in \citet{barnes16}.

Figure \ref{fig:Lbol_model} shows the bolometric light curves
for all the models.
Overall properties of the bolometric light curves are similar
to the simple models in Section \ref{sec:simple}.
The luminosity of each model at a few days after the merger is, however, 
higher than that of the simple model because the thermalization
factor is higher than 0.25 at the early time.
At $\gsim 5$ days after the merger, the luminosities decline faster
than those in the simple models due to decreasing thermalization efficiency.
This trend is most notable for the case of $\Ye = 0.3$
because of the smallest amounts of the second peak elements
with $Z \sim$ 50 (see Figure \ref{fig:abundance})
that have dominant contributions to the radioactive heating
\citep{metzger10,wanajo14}.
At late times ($> 14$ days) when the heating from fission becomes
important \citep{wanajo14,hotokezaka16},
the dynamical ejecta model (of $\Ye = 0.1-0.4$) that contains
the transuranic elements gives the highest luminosity.

Figures \ref{fig:mag} and \ref{fig:spec} show
  the multi-color light curves and
  spectra for three models, respectively.
The spectra of the dynamical ejecta model
(APR4-1215) are very red, which peaks in near infrared 
at $t = 1-20$ days.
On the other hand, the post-merger ejecta model with $\Ye = 0.3$
has a peak in optical at $t \lsim 5$ days.
As a result, the post-merger ejecta model with $\Ye = 0.3$
is much brighter than the dynamical ejecta model in optical,
especially in $u$, $g$, and $r$ bands.

The properties of the light curves of the post-merger ejecta model
with $\Ye = 0.25$ are in between the other two models,
as expected from the intermediate opacities.
Therefore, this model has hybrid properties;
the optical brightness is higher than that of dynamical ejecta model
and the near-infrared brightness is not as faint as that of
the post-merger ejecta with $\Ye = 0.3$ (Figure \ref{fig:spec}).

Our results confirm the presence of ``blue kilonova''
that was previously suggested based on the use of iron opacity
for the light $r$-process elements \citep{metzger14,kasen15}.
For 0.01 $\Msun$ of Lanthanide-free ($\Ye$ = 0.3) ejecta,
the optical brightness reaches the absolute magnitude of
$M = -14$ mag in $g$ and $r$ bands within a few days after the merger.
This corresponds to 21.0 mag and 22.5 mag at 100 Mpc and 200 Mpc, respectively.
Thanks to the relatively blue color,
this emission is detectable with
1m-class and 2m-class telescopes, respectively.

It should be noted that the observability of blue kilonova
from Lanthanide-free post-merger ejecta depends on
the properties of preceding dynamical ejecta as discussed in
\citet{kasen15}.
If Lanthanide-rich dynamical ejecta are present in all the direction,
the blue kilonova emission is likely to be absorbed.
However, recent relativistic simulations with neutrino interaction show that
dynamical ejecta can have relatively high $\Ye$ 
near the polar regions \citep[see, \eg][]{sekiguchi15,radice16,foucart16}.
In such case, the blue emission from the post-merger ejecta can be
observable from the polar direction without being absorbed.
To test this hypothesis, it is necessary to
consistently model the dynamical and post-merger ejecta.
It is also noted that our simulations cannot predict the emission
within $\sim 1$ day after the merger
due to lack of the atomic data of more ionized elements.
Emission at such early times can peak at optical or
even ultraviolet wavelengths \citep{metzger15,gottlieb17},
and therefore, it will also be a good target for
follow-up observations especially with small telescopes.

\section{Summary}
\label{sec:summary}

We have newly performed atomic structure calculations
for Se ($Z = 34$), Ru ($Z = 44$), Te ($Z = 52$), Ba ($Z = 56$),
Nd ($Z = 60$), and Er ($Z = 68$)
to construct the atomic data for a wide range of $r$-process elements.
By using two different atomic codes,
we confirmed that the atomic structure calculations
gave uncertainties in opacities by only a factor of up to about 2.
We found that the opacities from the bound-bound transitions
of open f-shell elements were the highest
from ultraviolet to near-infrared wavelengths,
while those of open s-shell, d-shell, and p-shell elements
were lower and concentrated in ultraviolet and optical wavelengths.

Using our new atomic data, we performed multi-wavelength
radiative transfer simulations
to predict a possible variety of kilonova emission.
We found that, even for the same ejecta mass,
the optical brightness varied by $> 2$ mag
depending on the distribution of elemental abundances.
If the blue emission from the post-merger, Lanthanide-free ejecta with 0.01 $\Msun$
is observable without being absorbed by preceding dynamical ejecta,
the brightness will reach the absolute magnitude of $M = -14$ mag in $g$ and $r$ bands
within a few days after the merger.
This corresponds to 21.0 mag and 22.5 mag at 100 Mpc and 200 Mpc,
which is detectable with 1m-class and 2m-class telescopes, respectively.

\acknowledgments
We thank Kenta Hotokezaka, Masaru Shibata, Nobuya Nishimura, Kenta Kiuchi,
and Koutarou Kyutoku for providing results of simulations
and fruitful discussion.
MT thanks the Institute for Nuclear Theory (INT)
at the University of Washington
for its hospitality and the Department of Energy for
partial support during the completion of this work.
MT also thanks Rodrigo Fern{\'a}ndez, Brian Metzger, Daniel Kasen,
and Gabriel Martinez-Pinedo for organizing the workshop
and providing the nice research environment at INT.
Numerical simulations presented in this paper 
were carried out with Cray XC30 at Center for Computational Astrophysics,
National Astronomical Observatory of Japan.
Computations by GG were performed on resources at the High Performance
Computing Center ``HPC Sauletekis'' of the Faculty of Physics
at Vilnius University.
This research was supported by
the NINS program for cross-disciplinary science study,
Inoue Science Research Award from Inoue Foundation for Science,
the RIKEN iTHES project,
a post-K computer project (Priority issue No. 9) of MEXT,
and the Grant-in-Aid for Scientific Research from
JSPS (15H00782, 15H02075, 15K05077, 16H02183, 16H06341, 16K17706,
17K05391, 26400237) and MEXT (17H06357, 17H06363).


\begin{thebibliography}{83}
\expandafter\ifx\csname natexlab\endcsname\relax\def\natexlab#1{#1}\fi

\bibitem[{{Abbott} {et~al.}(2016{\natexlab{a}}){Abbott}, {Abbott}, {Abbott},
  {Abernathy}, {Acernese}, {Ackley}, {Adams}, {Adams}, {Addesso}, {Adhikari},
  \& et~al.}]{abbott16b}
{Abbott}, B.~P., {et~al.} 2016{\natexlab{a}}, Physical Review Letters, 116,
  241103

\bibitem[{{Abbott} {et~al.}(2016{\natexlab{b}}){Abbott}, {Abbott}, {Abbott},
  {Abernathy}, {Acernese}, {Ackley}, {Adams}, {Adams}, {Addesso}, {Adhikari},
  \& et~al.}]{abbott16followup}
---. 2016{\natexlab{b}}, \apjl, 826, L13

\bibitem[{{Abbott} {et~al.}(2016{\natexlab{c}}){Abbott}, {Abbott}, {Abbott},
  {Abernathy}, {Acernese}, {Ackley}, {Adams}, {Adams}, {Addesso}, {Adhikari},
  \& et~al.}]{abbott16}
---. 2016{\natexlab{c}}, Physical Review Letters, 116, 061102

\bibitem[{{Abbott} {et~al.}(2017){Abbott}, {Abbott}, {Abbott}, {Acernese},
  {Ackley}, {Adams}, {Adams}, {Addesso}, {Adhikari}, {Adya}, \&
  et~al.}]{abbott17}
---. 2017, Physical Review Letters, 118, 221101

\bibitem[{{Abbott} {et~al.}(2016{\natexlab{d}}){Abbott}, {LIGO Scientific
  Collaboration}, \& {Virgo Collaboration}}]{abbott16review}
{Abbott}, B.~P., {LIGO Scientific Collaboration}, \& {Virgo Collaboration}.
  2016{\natexlab{d}}, Living Reviews in Relativity, 19

\bibitem[{{Badnell}(2011)}]{badnell11}
{Badnell}, N.~R. 2011, Computer Physics Communications, 182, 1528

\bibitem[{{Bar-Shalom} {et~al.}(2001){Bar-Shalom}, {Klapisch}, \&
  {Oreg}}]{bar-shalom01}
{Bar-Shalom}, A., {Klapisch}, M., \& {Oreg}, J. 2001, \jqsrt, 71, 169

\bibitem[{{Barnes} \& {Kasen}(2013)}]{barnes13}
{Barnes}, J., \& {Kasen}, D. 2013, \apj, 775, 18

\bibitem[{{Barnes} {et~al.}(2016){Barnes}, {Kasen}, {Wu}, \&
  {Mart{\'{\i}}nez-Pinedo}}]{barnes16}
{Barnes}, J., {Kasen}, D., {Wu}, M.-R., \& {Mart{\'{\i}}nez-Pinedo}, G. 2016,
  \apj, 829, 110

\bibitem[{{Berger} {et~al.}(2013){Berger}, {Fong}, \& {Chornock}}]{berger13}
{Berger}, E., {Fong}, W., \& {Chornock}, R. 2013, \apjl, 774, L23

\bibitem[{{Ciolfi} {et~al.}(2017){Ciolfi}, {Kastaun}, {Giacomazzo}, {Endrizzi},
  {Siegel}, \& {Perna}}]{ciolfi17}
{Ciolfi}, R., {Kastaun}, W., {Giacomazzo}, B., {Endrizzi}, A., {Siegel}, D.~M.,
  \& {Perna}, R. 2017, \prd, 95, 063016

\bibitem[{{Cowperthwaite} {et~al.}(2016){Cowperthwaite}, {Berger},
  {Soares-Santos}, {Annis}, {Brout}, {Brown}, {Buckley-Geer}, {Cenko}, {Chen},
  {Chornock}, {Diehl}, {Doctor}, {Drlica-Wagner}, {Drout}, {Farr}, {Finley},
  {Foley}, {Fong}, {Fox}, {Frieman}, {Garcia-Bellido}, {Gill}, {Gruendl},
  {Herner}, {Holz}, {Kasen}, {Kessler}, {Lin}, {Margutti}, {Marriner},
  {Matheson}, {Metzger}, {Neilsen}, {Quataert}, {Rest}, {Sako}, {Scolnic},
  {Smith}, {Sobreira}, {Strampelli}, {Villar}, {Walker}, {Wester}, {Williams},
  {Yanny}, {Abbott}, {Abdalla}, {Allam}, {Armstrong}, {Bechtol},
  {Benoit-L{\'e}vy}, {Bertin}, {Brooks}, {Burke}, {Carnero Rosell}, {Carrasco
  Kind}, {Carretero}, {Castander}, {Cunha}, {D'Andrea}, {da Costa}, {Desai},
  {Dietrich}, {Evrard}, {Fausti Neto}, {Fosalba}, {Gerdes}, {Giannantonio},
  {Goldstein}, {Gruen}, {Gutierrez}, {Honscheid}, {James}, {Johnson},
  {Johnson}, {Krause}, {Kuehn}, {Kuropatkin}, {Lima}, {Maia}, {Marshall},
  {Menanteau}, {Miquel}, {Mohr}, {Nichol}, {Nord}, {Ogando}, {Plazas}, {Reil},
  {Romer}, {Sanchez}, {Scarpine}, {Sevilla-Noarbe}, {Smith}, {Suchyta},
  {Tarle}, {Thomas}, {Thomas}, {Tucker}, {Weller}, \& {DES
  Collaboration}}]{cowperthwaite16}
{Cowperthwaite}, P.~S., {et~al.} 2016, \apjl, 826, L29

\bibitem[{{Dessart} {et~al.}(2009){Dessart}, {Ott}, {Burrows}, {Rosswog}, \&
  {Livne}}]{dessart09}
{Dessart}, L., {Ott}, C.~D., {Burrows}, A., {Rosswog}, S., \& {Livne}, E. 2009,
  \apj, 690, 1681

\bibitem[{{Dyall} {et~al.}(1989){Dyall}, {Grant}, {Johnson}, {Parpia}, \&
  {Plummer}}]{dyall89}
{Dyall}, K.~G., {Grant}, I.~P., {Johnson}, C.~T., {Parpia}, F.~A., \&
  {Plummer}, E.~P. 1989, Computer Physics Communications, 55, 425

\bibitem[{{Eastman} \& {Pinto}(1993)}]{eastman93}
{Eastman}, R.~G., \& {Pinto}, P.~A. 1993, \apj, 412, 731

\bibitem[{{Fern{\'a}ndez} {et~al.}(2015{\natexlab{a}}){Fern{\'a}ndez}, {Kasen},
  {Metzger}, \& {Quataert}}]{fernandez15nsns}
{Fern{\'a}ndez}, R., {Kasen}, D., {Metzger}, B.~D., \& {Quataert}, E.
  2015{\natexlab{a}}, \mnras, 446, 750

\bibitem[{{Fern{\'a}ndez} \& {Metzger}(2013)}]{fernandez13}
{Fern{\'a}ndez}, R., \& {Metzger}, B.~D. 2013, \mnras, 435, 502

\bibitem[{{Fern{\'a}ndez} \& {Metzger}(2016)}]{fernandez16}
---. 2016, Annual Review of Nuclear and Particle Science, 66, 23

\bibitem[{{Fern{\'a}ndez} {et~al.}(2015{\natexlab{b}}){Fern{\'a}ndez},
  {Quataert}, {Schwab}, {Kasen}, \& {Rosswog}}]{fernandez15bhns}
{Fern{\'a}ndez}, R., {Quataert}, E., {Schwab}, J., {Kasen}, D., \& {Rosswog},
  S. 2015{\natexlab{b}}, \mnras, 449, 390

\bibitem[{{Fontes} {et~al.}(2017){Fontes}, {Fryer}, {Hungerford}, {Wollaeger},
  {Rosswog}, \& {Berger}}]{fontes17}
{Fontes}, C.~J., {Fryer}, C.~L., {Hungerford}, A.~L., {Wollaeger}, R.~T.,
  {Rosswog}, S., \& {Berger}, E. 2017, arXiv:1702.02990

\bibitem[{{Foucart} {et~al.}(2016){Foucart}, {O'Connor}, {Roberts}, {Kidder},
  {Pfeiffer}, \& {Scheel}}]{foucart16}
{Foucart}, F., {O'Connor}, E., {Roberts}, L., {Kidder}, L.~E., {Pfeiffer},
  H.~P., \& {Scheel}, M.~A. 2016, \prd, 94, 123016

\bibitem[{{Froese Fischer} {et~al.}(2016){Froese Fischer}, {Godefroid},
  {Brage}, {J{\"o}nsson}, \& {Gaigalas}}]{froesefischer16}
{Froese Fischer}, C., {Godefroid}, M., {Brage}, T., {J{\"o}nsson}, P., \&
  {Gaigalas}, G. 2016, Journal of Physics B Atomic Molecular Physics, 49,
  182004

\bibitem[{{Fujibayashi} {et~al.}(2017){Fujibayashi}, {Sekiguchi}, {Kiuchi}, \&
  {Shibata}}]{fujibayashi17}
{Fujibayashi}, S., {Sekiguchi}, Y., {Kiuchi}, K., \& {Shibata}, M. 2017,
  arXiv:1703.10191

\bibitem[{{Gaigalas} {et~al.}(2017){Gaigalas}, {Fischer}, {Rynkun}, \&
  {J{\"o}nsson}}]{gaigalas17}
{Gaigalas}, G., {Fischer}, C., {Rynkun}, P., \& {J{\"o}nsson}, P. 2017, Atoms,
  5, 6

\bibitem[{{Giacomazzo} {et~al.}(2015){Giacomazzo}, {Zrake}, {Duffell},
  {MacFadyen}, \& {Perna}}]{giacomazzo15}
{Giacomazzo}, B., {Zrake}, J., {Duffell}, P.~C., {MacFadyen}, A.~I., \&
  {Perna}, R. 2015, \apj, 809, 39

\bibitem[{{Goriely} {et~al.}(2015){Goriely}, {Bauswein}, {Just}, {Pllumbi}, \&
  {Janka}}]{goriely15}
{Goriely}, S., {Bauswein}, A., {Just}, O., {Pllumbi}, E., \& {Janka}, H.-T.
  2015, \mnras, 452, 3894

\bibitem[{{Gottlieb} {et~al.}(2017){Gottlieb}, {Nakar}, \&
  {Piran}}]{gottlieb17}
{Gottlieb}, O., {Nakar}, E., \& {Piran}, T. 2017, arXiv:1705.10797

\bibitem[{{Grant}(2007)}]{grant07}
{Grant}, I.~P. 2007, {Relativistic Quantum Theory of Atoms and Molecules}
  (Springer Science+Business Media)

\bibitem[{{Hotokezaka} {et~al.}(2013{\natexlab{a}}){Hotokezaka}, {Kiuchi},
  {Kyutoku}, {Okawa}, {Sekiguchi}, {Shibata}, \& {Taniguchi}}]{hotokezaka13}
{Hotokezaka}, K., {Kiuchi}, K., {Kyutoku}, K., {Okawa}, H., {Sekiguchi}, Y.,
  {Shibata}, M., \& {Taniguchi}, K. 2013{\natexlab{a}}, \prd, 87, 024001

\bibitem[{{Hotokezaka} {et~al.}(2013{\natexlab{b}}){Hotokezaka}, {Kyutoku},
  {Tanaka}, {Kiuchi}, {Sekiguchi}, {Shibata}, \& {Wanajo}}]{hotokezaka13b}
{Hotokezaka}, K., {Kyutoku}, K., {Tanaka}, M., {Kiuchi}, K., {Sekiguchi}, Y.,
  {Shibata}, M., \& {Wanajo}, S. 2013{\natexlab{b}}, \apjl, 778, L16

\bibitem[{{Hotokezaka} {et~al.}(2016){Hotokezaka}, {Wanajo}, {Tanaka}, {Bamba},
  {Terada}, \& {Piran}}]{hotokezaka16}
{Hotokezaka}, K., {Wanajo}, S., {Tanaka}, M., {Bamba}, A., {Terada}, Y., \&
  {Piran}, T. 2016, \mnras, 459, 35

\bibitem[{{Jin} {et~al.}(2016){Jin}, {Hotokezaka}, {Li}, {Tanaka}, {D'Avanzo},
  {Fan}, {Covino}, {Wei}, \& {Piran}}]{jin16}
{Jin}, Z.-P., {et~al.} 2016, Nature Communications, 7, 12898

\bibitem[{{Jin} {et~al.}(2015){Jin}, {Li}, {Cano}, {Covino}, {Fan}, \&
  {Wei}}]{jin15}
{Jin}, Z.-P., {Li}, X., {Cano}, Z., {Covino}, S., {Fan}, Y.-Z., \& {Wei}, D.-M.
  2015, \apjl, 811, L22

\bibitem[{{J{\"o}nsson} {et~al.}(2013){J{\"o}nsson}, {Gaigalas}, {Biero{\'n}},
  {Fischer}, \& {Grant}}]{jonsson13}
{J{\"o}nsson}, P., {Gaigalas}, G., {Biero{\'n}}, J., {Fischer}, C.~F., \&
  {Grant}, I.~P. 2013, Computer Physics Communications, 184, 2197

\bibitem[{{Just} {et~al.}(2015){Just}, {Bauswein}, {Pulpillo}, {Goriely}, \&
  {Janka}}]{just15}
{Just}, O., {Bauswein}, A., {Pulpillo}, R.~A., {Goriely}, S., \& {Janka}, H.-T.
  2015, \mnras, 448, 541

\bibitem[{{Karp} {et~al.}(1977){Karp}, {Lasher}, {Chan}, \&
  {Salpeter}}]{karp77}
{Karp}, A.~H., {Lasher}, G., {Chan}, K.~L., \& {Salpeter}, E.~E. 1977, \apj,
  214, 161

\bibitem[{{Kasen} {et~al.}(2013){Kasen}, {Badnell}, \& {Barnes}}]{kasen13}
{Kasen}, D., {Badnell}, N.~R., \& {Barnes}, J. 2013, \apj, 774, 25

\bibitem[{{Kasen} {et~al.}(2015){Kasen}, {Fern{\'a}ndez}, \&
  {Metzger}}]{kasen15}
{Kasen}, D., {Fern{\'a}ndez}, R., \& {Metzger}, B.~D. 2015, \mnras, 450, 1777

\bibitem[{{Kasen} {et~al.}(2006){Kasen}, {Thomas}, \& {Nugent}}]{kasen06}
{Kasen}, D., {Thomas}, R.~C., \& {Nugent}, P. 2006, \apj, 651, 366

\bibitem[{{Kasliwal} {et~al.}(2016){Kasliwal}, {Cenko}, {Singer}, {Corsi},
  {Cao}, {Barlow}, {Bhalerao}, {Bellm}, {Cook}, {Duggan}, {Ferretti}, {Frail},
  {Horesh}, {Kendrick}, {Kulkarni}, {Lunnan}, {Palliyaguru}, {Laher}, {Masci},
  {Manulis}, {Miller}, {Nugent}, {Perley}, {Prince}, {Quimby}, {Rana},
  {Rebbapragada}, {Sesar}, {Singhal}, {Surace}, \& {Van Sistine}}]{kasliwal16}
{Kasliwal}, M.~M., {et~al.} 2016, \apjl, 824, L24

\bibitem[{{Kiuchi} {et~al.}(2014){Kiuchi}, {Kyutoku}, {Sekiguchi}, {Shibata},
  \& {Wada}}]{kiuchi14}
{Kiuchi}, K., {Kyutoku}, K., {Sekiguchi}, Y., {Shibata}, M., \& {Wada}, T.
  2014, \prd, 90, 041502

\bibitem[{{Kiuchi} {et~al.}(2015){Kiuchi}, {Sekiguchi}, {Kyutoku}, {Shibata},
  {Taniguchi}, \& {Wada}}]{kiuchi15}
{Kiuchi}, K., {Sekiguchi}, Y., {Kyutoku}, K., {Shibata}, M., {Taniguchi}, K.,
  \& {Wada}, T. 2015, \prd, 92, 064034

\bibitem[{Kramida {et~al.}(2015)Kramida, {Yu.~Ralchenko}, Reader, \& {and NIST
  ASD Team}}]{kramida15}
Kramida, A., {Yu.~Ralchenko}, Reader, J., \& {and NIST ASD Team}. 2015, {NIST
  Atomic Spectra Database (ver. 5.3) {\tt{http://physics.nist.gov/asd}}.
  National Institute of Standards and Technology, Gaithersburg, MD.}

\bibitem[{{Kulkarni}(2005)}]{kulkarni05}
{Kulkarni}, S.~R. 2005, arXiv:astro-ph/0510256

\bibitem[{{Kupka} {et~al.}(1999){Kupka}, {Piskunov}, {Ryabchikova}, {Stempels},
  \& {Weiss}}]{kupka99}
{Kupka}, F., {Piskunov}, N., {Ryabchikova}, T.~A., {Stempels}, H.~C., \&
  {Weiss}, W.~W. 1999, \aaps, 138, 119

\bibitem[{{Kupka} {et~al.}(2000){Kupka}, {Ryabchikova}, {Piskunov}, {Stempels},
  \& {Weiss}}]{kupka00}
{Kupka}, F.~G., {Ryabchikova}, T.~A., {Piskunov}, N.~E., {Stempels}, H.~C., \&
  {Weiss}, W.~W. 2000, Baltic Astronomy, 9, 590

\bibitem[{{Kurucz} \& {Bell}(1995)}]{kurucz95}
{Kurucz}, R., \& {Bell}, B. 1995, Atomic Line Data (R.L.~Kurucz and B.~Bell)
  Kurucz CD-ROM No.~23.~Cambridge, Mass.: Smithsonian Astrophysical
  Observatory, 1995., 23

\bibitem[{{Lehner} {et~al.}(2016){Lehner}, {Liebling}, {Palenzuela},
  {Caballero}, {O'Connor}, {Anderson}, \& {Neilsen}}]{lehner16}
{Lehner}, L., {Liebling}, S.~L., {Palenzuela}, C., {Caballero}, O.~L.,
  {O'Connor}, E., {Anderson}, M., \& {Neilsen}, D. 2016, Classical and Quantum
  Gravity, 33, 184002

\bibitem[{{Li} \& {Paczy{\'n}ski}(1998)}]{li98}
{Li}, L.-X., \& {Paczy{\'n}ski}, B. 1998, \apjl, 507, L59

\bibitem[{{Lippuner} {et~al.}(2017){Lippuner}, {Fern{\'a}ndez}, {Roberts},
  {Foucart}, {Kasen}, {Metzger}, \& {Ott}}]{lippuner17}
{Lippuner}, J., {Fern{\'a}ndez}, R., {Roberts}, L.~F., {Foucart}, F., {Kasen},
  D., {Metzger}, B.~D., \& {Ott}, C.~D. 2017, arXiv:1703.06216

\bibitem[{{Martin} {et~al.}(2015){Martin}, {Perego}, {Arcones}, {Thielemann},
  {Korobkin}, \& {Rosswog}}]{martin15}
{Martin}, D., {Perego}, A., {Arcones}, A., {Thielemann}, F.-K., {Korobkin}, O.,
  \& {Rosswog}, S. 2015, \apj, 813, 2

\bibitem[{{Metzger}(2017)}]{metzger17}
{Metzger}, B.~D. 2017, Living Reviews in Relativity, 20, 3

\bibitem[{{Metzger} {et~al.}(2015){Metzger}, {Bauswein}, {Goriely}, \&
  {Kasen}}]{metzger15}
{Metzger}, B.~D., {Bauswein}, A., {Goriely}, S., \& {Kasen}, D. 2015, \mnras,
  446, 1115

\bibitem[{{Metzger} \& {Berger}(2012)}]{metzger12}
{Metzger}, B.~D., \& {Berger}, E. 2012, \apj, 746, 48

\bibitem[{{Metzger} \& {Fern{\'a}ndez}(2014)}]{metzger14}
{Metzger}, B.~D., \& {Fern{\'a}ndez}, R. 2014, \mnras, 441, 3444

\bibitem[{{Metzger} {et~al.}(2010){Metzger}, {Mart{\'{\i}}nez-Pinedo},
  {Darbha}, {Quataert}, {Arcones}, {Kasen}, {Thomas}, {Nugent}, {Panov}, \&
  {Zinner}}]{metzger10}
{Metzger}, B.~D., {et~al.} 2010, \mnras, 406, 2650

\bibitem[{{Morokuma} {et~al.}(2016){Morokuma}, {Tanaka}, {Asakura}, {Abe},
  {Tristram}, {Utsumi}, {Doi}, {Fujisawa}, {Itoh}, {Itoh}, {Kawabata}, {Kawai},
  {Kuroda}, {Matsubayashi}, {Motohara}, {Murata}, {Nagayama}, {Ohta}, {Saito},
  {Tamura}, {Tominaga}, {Uemura}, {Yanagisawa}, {Yatsu}, \&
  {Yoshida}}]{morokuma16}
{Morokuma}, T., {et~al.} 2016, \pasj, 68, L9

\bibitem[{{Olsen} {et~al.}(1995){Olsen}, {Godefroid}, {J{\"o}nsson},
  {Malmqvist}, \& {Fischer}}]{olsen95}
{Olsen}, J., {Godefroid}, M.~R., {J{\"o}nsson}, P., {Malmqvist}, P.~{\AA}., \&
  {Fischer}, C.~F. 1995, \pre, 52, 4499

\bibitem[{{Perego} {et~al.}(2014){Perego}, {Rosswog}, {Cabez{\'o}n},
  {Korobkin}, {K{\"a}ppeli}, {Arcones}, \& {Liebend{\"o}rfer}}]{perego14}
{Perego}, A., {Rosswog}, S., {Cabez{\'o}n}, R.~M., {Korobkin}, O.,
  {K{\"a}ppeli}, R., {Arcones}, A., \& {Liebend{\"o}rfer}, M. 2014, \mnras,
  443, 3134

\bibitem[{{Piran} {et~al.}(2014){Piran}, {Korobkin}, \& {Rosswog}}]{piran14}
{Piran}, T., {Korobkin}, O., \& {Rosswog}, S. 2014, arXiv:1401.2166

\bibitem[{{Piskunov} {et~al.}(1995){Piskunov}, {Kupka}, {Ryabchikova}, {Weiss},
  \& {Jeffery}}]{piskunov95}
{Piskunov}, N.~E., {Kupka}, F., {Ryabchikova}, T.~A., {Weiss}, W.~W., \&
  {Jeffery}, C.~S. 1995, \aaps, 112, 525

\bibitem[{{Price} \& {Rosswog}(2006)}]{price06}
{Price}, D.~J., \& {Rosswog}, S. 2006, Science, 312, 719

\bibitem[{{Radice} {et~al.}(2016){Radice}, {Galeazzi}, {Lippuner}, {Roberts},
  {Ott}, \& {Rezzolla}}]{radice16}
{Radice}, D., {Galeazzi}, F., {Lippuner}, J., {Roberts}, L.~F., {Ott}, C.~D.,
  \& {Rezzolla}, L. 2016, \mnras, 460, 3255

\bibitem[{{Rosswog}(2015)}]{rosswog15}
{Rosswog}, S. 2015, International Journal of Modern Physics D, 24, 1530012

\bibitem[{{Rosswog} {et~al.}(2017){Rosswog}, {Feindt}, {Korobkin}, {Wu},
  {Sollerman}, {Goobar}, \& {Martinez-Pinedo}}]{rosswog17}
{Rosswog}, S., {Feindt}, U., {Korobkin}, O., {Wu}, M.-R., {Sollerman}, J.,
  {Goobar}, A., \& {Martinez-Pinedo}, G. 2017, Classical and Quantum Gravity,
  34, 104001

\bibitem[{{Ryabchikova} {et~al.}(1997){Ryabchikova}, {Piskunov}, {Kupka}, \&
  {Weiss}}]{ryabchikova97}
{Ryabchikova}, T.~A., {Piskunov}, N.~E., {Kupka}, F., \& {Weiss}, W.~W. 1997,
  Baltic Astronomy, 6, 244

\bibitem[{{Sekiguchi} {et~al.}(2015){Sekiguchi}, {Kiuchi}, {Kyutoku}, \&
  {Shibata}}]{sekiguchi15}
{Sekiguchi}, Y., {Kiuchi}, K., {Kyutoku}, K., \& {Shibata}, M. 2015, \prd, 91,
  064059

\bibitem[{{Sekiguchi} {et~al.}(2016){Sekiguchi}, {Kiuchi}, {Kyutoku},
  {Shibata}, \& {Taniguchi}}]{sekiguchi16}
{Sekiguchi}, Y., {Kiuchi}, K., {Kyutoku}, K., {Shibata}, M., \& {Taniguchi}, K.
  2016, \prd, 93, 124046

\bibitem[{{Shibata} {et~al.}(2017){Shibata}, {Kiuchi}, \&
  {Sekiguchi}}]{shibata17}
{Shibata}, M., {Kiuchi}, K., \& {Sekiguchi}, Y. 2017, \prd, 95, 083005

\bibitem[{{Siegel} \& {Metzger}(2017)}]{siegel17}
{Siegel}, D.~M., \& {Metzger}, B.~D. 2017, arXiv:1705.05473

\bibitem[{{Simmerer} {et~al.}(2004){Simmerer}, {Sneden}, {Cowan}, {Collier},
  {Woolf}, \& {Lawler}}]{simmerer04}
{Simmerer}, J., {Sneden}, C., {Cowan}, J.~J., {Collier}, J., {Woolf}, V.~M., \&
  {Lawler}, J.~E. 2004, \apj, 617, 1091

\bibitem[{{Smartt} {et~al.}(2016){Smartt}, {Chambers}, {Smith}, {Huber},
  {Young}, {Cappellaro}, {Wright}, {Coughlin}, {Schultz}, {Denneau},
  {Flewelling}, {Heinze}, {Magnier}, {Primak}, {Rest}, {Sherstyuk}, {Stalder},
  {Stubbs}, {Tonry}, {Waters}, {Willman}, {Anderson}, {Baltay}, {Botticella},
  {Campbell}, {Dennefeld}, {Chen}, {Della Valle}, {Elias-Rosa}, {Fraser},
  {Inserra}, {Kankare}, {Kotak}, {Kupfer}, {Harmanen}, {Galbany}, {Gal-Yam},
  {Le Guillou}, {Lyman}, {Maguire}, {Mitra}, {Nicholl}, {Olivares E},
  {Rabinowitz}, {Razza}, {Sollerman}, {Smith}, {Terreran}, {Valenti}, {Gibson},
  \& {Goggia}}]{smartt16}
{Smartt}, S.~J., {et~al.} 2016, \mnras, 462, 4094

\bibitem[{{Soares-Santos} {et~al.}(2016){Soares-Santos}, {Kessler}, {Berger},
  {Annis}, {Brout}, {Buckley-Geer}, {Chen}, {Cowperthwaite}, {Diehl}, {Doctor},
  {Drlica-Wagner}, {Farr}, {Finley}, {Flaugher}, {Foley}, {Frieman}, {Gruendl},
  {Herner}, {Holz}, {Lin}, {Marriner}, {Neilsen}, {Rest}, {Sako}, {Scolnic},
  {Sobreira}, {Walker}, {Wester}, {Yanny}, {Abbott}, {Abdalla}, {Allam},
  {Armstrong}, {Banerji}, {Benoit-L{\'e}vy}, {Bernstein}, {Bertin}, {Brown},
  {Burke}, {Capozzi}, {Carnero Rosell}, {Carrasco Kind}, {Carretero},
  {Castander}, {Cenko}, {Chornock}, {Crocce}, {D'Andrea}, {da Costa}, {Desai},
  {Dietrich}, {Drout}, {Eifler}, {Estrada}, {Evrard}, {Fairhurst}, {Fernandez},
  {Fischer}, {Fong}, {Fosalba}, {Fox}, {Fryer}, {Garcia-Bellido}, {Gaztanaga},
  {Gerdes}, {Goldstein}, {Gruen}, {Gutierrez}, {Honscheid}, {James},
  {Karliner}, {Kasen}, {Kent}, {Kuropatkin}, {Kuehn}, {Lahav}, {Li}, {Lima},
  {Maia}, {Margutti}, {Martini}, {Matheson}, {McMahon}, {Metzger}, {Miller},
  {Miquel}, {Mohr}, {Nichol}, {Nord}, {Ogando}, {Peoples}, {Plazas},
  {Quataert}, {Romer}, {Roodman}, {Rykoff}, {Sanchez}, {Scarpine}, {Schindler},
  {Schubnell}, {Sevilla-Noarbe}, {Sheldon}, {Smith}, {Smith}, {Smith},
  {Stebbins}, {Sutton}, {Swanson}, {Tarle}, {Thaler}, {Thomas}, {Tucker},
  {Vikram}, {Wechsler}, {Weller}, \& {DES Collaboration}}]{soares-santos16}
{Soares-Santos}, M., {et~al.} 2016, \apjl, 823, L33

\bibitem[{{Tanaka}(2016)}]{tanaka16}
{Tanaka}, M. 2016, Advances in Astronomy, 2016, 634197

\bibitem[{{Tanaka} \& {Hotokezaka}(2013)}]{tanaka13}
{Tanaka}, M., \& {Hotokezaka}, K. 2013, \apj, 775, 113

\bibitem[{{Tanaka} {et~al.}(2014){Tanaka}, {Hotokezaka}, {Kyutoku}, {Wanajo},
  {Kiuchi}, {Sekiguchi}, \& {Shibata}}]{tanaka14}
{Tanaka}, M., {Hotokezaka}, K., {Kyutoku}, K., {Wanajo}, S., {Kiuchi}, K.,
  {Sekiguchi}, Y., \& {Shibata}, M. 2014, \apj, 780, 31

\bibitem[{{Tanvir} {et~al.}(2013){Tanvir}, {Levan}, {Fruchter}, {Hjorth},
  {Hounsell}, {Wiersema}, \& {Tunnicliffe}}]{tanvir13}
{Tanvir}, N.~R., {Levan}, A.~J., {Fruchter}, A.~S., {Hjorth}, J., {Hounsell},
  R.~A., {Wiersema}, K., \& {Tunnicliffe}, R.~L. 2013, \nat, 500, 547

\bibitem[{{Wanajo} \& {Janka}(2012)}]{wanajo12}
{Wanajo}, S., \& {Janka}, H.-T. 2012, \apj, 746, 180

\bibitem[{{Wanajo} {et~al.}(2014){Wanajo}, {Sekiguchi}, {Nishimura}, {Kiuchi},
  {Kyutoku}, \& {Shibata}}]{wanajo14}
{Wanajo}, S., {Sekiguchi}, Y., {Nishimura}, N., {Kiuchi}, K., {Kyutoku}, K., \&
  {Shibata}, M. 2014, \apjl, 789, L39

\bibitem[{{Wollaeger} {et~al.}(2017){Wollaeger}, {Korobkin}, {Fontes},
  {Rosswog}, {Even}, {Fryer}, {Sollerman}, {Hungerford}, {van Rossum}, \&
  {Wollaber}}]{wollaeger17}
{Wollaeger}, R.~T., {et~al.} 2017, arXiv:1705.07084

\bibitem[{{Wu} {et~al.}(2016){Wu}, {Fern{\'a}ndez}, {Mart{\'{\i}}nez-Pinedo},
  \& {Metzger}}]{wu16}
{Wu}, M.-R., {Fern{\'a}ndez}, R., {Mart{\'{\i}}nez-Pinedo}, G., \& {Metzger},
  B.~D. 2016, \mnras, 463, 2323

\bibitem[{{Yang} {et~al.}(2015){Yang}, {Jin}, {Li}, {Covino}, {Zheng},
  {Hotokezaka}, {Fan}, {Piran}, \& {Wei}}]{yang15}
{Yang}, B., {et~al.} 2015, Nature Communications, 6, 7323

\bibitem[{{Yoshida} {et~al.}(2017){Yoshida}, {Utsumi}, {Tominaga}, {Morokuma},
  {Tanaka}, {Asakura}, {Matsubayashi}, {Ohta}, {Abe}, {Chimasu}, {Furusawa},
  {Itoh}, {Itoh}, {Kanda}, {Kawabata}, {Kawabata}, {Koshida}, {Koshimoto},
  {Kuroda}, {Moritani}, {Motohara}, {Murata}, {Nagayama}, {Nakaoka}, {Nakata},
  {Nishioka}, {Saito}, {Terai}, {Tristram}, {Yanagisawa}, {Yasuda}, {Doi},
  {Fujisawa}, {Kawachi}, {Kawai}, {Tamura}, {Uemura}, \& {Yatsu}}]{yoshida17}
{Yoshida}, M., {et~al.} 2017, \pasj, 69, 9

\end{thebibliography}

\end{document}